\documentstyle[aaspp4,aabib]{article}
\newcommand{\kms}{km~s{$^{-1}$}}

\input epsf

\begin{document}

\title{ Synthetic Molecular Clouds from Supersonic MHD
        and Non-LTE Radiative Transfer Calculations}

\author{Paolo Padoan}
\affil{Theoretical Astrophysics Center,
       Juliane Maries Vej 30, DK-2100 Copenhagen,
       Denmark}
\author{Mika Juvela}
\affil{Helsinki University Observatory,
       T\"ahtitorninm\"aki, P.O.Box 14,
       SF-00014 University of Helsinki, Finland}
\author{John Bally}
\affil{Department of Astrophysics, Planetary, and Atmospheric Sciences,\\
       Center for Astrophysics and Space Astronomy,\\
       Campus Box 389, University of Colorado, Boulder CO 80309}
\author{\AA ke Nordlund}
\affil{Astronomical Observatory and Theoretical Astrophysics Center, \\
       Juliane Maries Vej 30, DK-2100 Copenhagen, Denmark}

\begin{abstract}

The dynamics of molecular clouds is characterized by supersonic
random motions in the presence of a magnetic field. We study
this situation using numerical solutions of the three--dimensional
compressible magneto--hydrodynamic (MHD) equations in a regime of
highly supersonic random motions.  The non-LTE radiative transfer 
calculations are performed through the complex density and velocity fields
obtained as solutions of the MHD equations, and more
than $5 \times 10^5$ spectra of $^{12}$CO, $^{13}$CO and CS are obtained. 
In this way we build synthetic molecular 
clouds of 5~pc and 20~pc diameter, evolved for about one dynamical 
time from their initial configuration. We use a numerical flow 
without gravity or external forcing. The flow is super--Alfv\'{e}nic 
and corresponds to model A of Padoan and Nordlund (1997). 
\nocite{Padoan+Nordlund97dyn}

Synthetic data consist of sets of 90$\times$90 
synthetic spectra with 60 velocity channels, in five 
molecular transitions: J=1$\rightarrow$0 and J=2$\rightarrow$1 
for $^{12}$CO and $^{13}$CO, and J=1$\rightarrow$0 for CS. 
Though we do not consider the effects of stellar radiation,
gravity, or mechanical energy input from discrete sources,
our models do contain the basic physics of magneto--fluid dynamics and
non LTE radiation transfer and are therefore more realistic than
previous calculations.  As a result, these synthetic maps and spectra 
bear a remarkable resemblance to the corresponding observations of 
real clouds.

\end{abstract}

\keywords{
turbulence - ISM: kinematics and dynamics- magnetic field }

\section{Introduction}

Stars form from the gravitational collapse of cores in molecular 
clouds. The conditions under which such collapse occurs are 
poorly understood.  Observations show that molecular clouds in which stars 
form are chaotic in nature.  An understanding 
of the initial conditions that lead to star formation must take into 
consideration the complex non--linear magneto--hydrodynamic behavior 
of these clouds.  Such studies are essential to an understanding of 
star formation on large scales; a process that is fundamental to the 
description of galaxy formation and the conversion of matter into 
light in the Universe.

Observations have shown that molecular clouds exhibit highly supersonic
line-widths and complex internal structure consisting
of clumps, sheets, and filaments (cf. Bally et al 1987).  
\nocite{Bally+87}
The persistence of supersonic motions has led to the postulate that
strong magnetic fields threat the clouds and prevent the efficient 
dissipation of these motions.  Therefore, the observed line widths
are assumed to be close to the Alfv\'{e}n speed in the cloud. 
However, there is little direct observational evidence for pervasive
magnetic fields of the required strength.
Furthermore, it has been shown that Alfv\'{e}n waves
do not not increase the dissipation time--scale significantly
(Zweibel \& Josafatsson 1983; Elmegreen 1985; Padoan \& Nordlund 1997), 
\nocite{Zweibel+Josafatsson83}
\nocite{Elmegreen85}
which was the original motivation for models of clouds supported 
by Alfv\'{e}n waves.

Other workers have proposed that the chaotic structure of molecular clouds and the
power law distribution of mass spectra of substructures is an indication
of turbulent dynamics. Possible manifestations of turbulent motions
are the Larson relations (Larson 1981), the hierarchical 
\nocite{Larson81}
structure of the density field (Scalo 1985), the observed intermittency 
\nocite{Scalo85}
(high velocity tails) in the line profiles (Falgarone \& Phillips 1990),
\nocite{Falgarone+Phillips90}
and the self--similarity of molecular cloud edges or mass spectra 
(Falgarone, Phillips \& Walker 1991; Elmegreen \& Falgarone 1996;
Falgarone \& Phillips 1996). 
\nocite{Falgarone+91,Elmegreen+Falgarone96,Falgarone+Phillips96}

Models by Padoan, Jones, and Nordlund (1997a) and Padoan and Nordlund (1997)
\nocite{Padoan+97ext,Padoan+Nordlund97dyn}
have shown that supersonic and super--Alfv\'{e}nic random flow models  
reproduce many aspects of the observed structure of molecular clouds,
such as for example the 
near--infrared stellar extinction determinations of Lada et al.\ (1994).
\nocite{Lada+94}
Moreover, Padoan and Nordlund (1997) have shown that these models are
consistent with the OH Zeeman measurements and constraints.  As we show
in this paper, these models also result in synthetic molecular cloud
spectra and maps that closely resemble real data.

Interpretation of observational data on molecular clouds requires assumptions about
the structure of the emitting gas (e.g.\ Zuckerman \& Evans 1974; 
\nocite{Zuckerman+Evans74}
Leung \& Liszt 1976; Baker 1976; Dickman 1978; Kwan \& Sanders 1986; 
Albrecht \& Kegel 1987; Tauber \& Goldsmith 1990; Tauber, 
Goldsmith \& Dickman 1991; Wolfire, Hollenbach \& Tielens 1993; 
Robert \& Pagani 1993; Park \& Hong 1995; Park, Hong \& Minh 1996; 
Juvela 1997).  In general, the assumptions used by these authors 
\nocite{Leung+Liszt76,Baker76,Dickman78,Kwan+Sanders86}
\nocite{Albrecht+Kegel87,Tauber+Goldsmith90,Tauber+91}
\nocite{Wolfire+93,Robert+Pagani93,Park+Hong95,Park+96,Juvela97}
are not consistent with the dynamics of molecular clouds, and are in most cases limited 
to the statement that molecular clouds are clumpy.  Stenholm \& Pudritz (1993) and 
\nocite{Stenholm+Pudritz93}
Falgarone et al. (1994) calculated synthetic molecular spectra from fluid 
\nocite{Falgarone+94}
models of clouds.  Stenholm \& Pudritz (1993) used a sticky particles code 
with an imposed spectrum of Alfv\'{e}n waves (Carlberg \& Pudritz 1990). 
\nocite{Carlberg+Pudritz90}
They calculated line profiles under the simple assumption of LTE. 
Falgarone et al. (1994) did not solve the radiative transfer problem but
calculated density weighted radial velocity profiles on the 
basis of the results of a turbulence simulation by Porter, 
Pouquet \& Woodward (1994).
\nocite{Porter+94}

A lot of work has been done to understand the clumpiness of molecular clouds, and
many observational studies contain lists of molecular clumps,
or cores, with their properties (size, mass, velocity dispersion, etc.). 
These studies have led to the discovery of scaling laws in the 
ISM, such as the molecular line width-size relation 
(Larson 1981; Leung, Kutner \& Mead 1982; Myers 1983; Quiroga 1983; 
Sanders, Scoville \& Solomon 1985; Crovisier, Dickey \& Kaz\`{e}s 1985; 
Goldsmith \& Arquilla 1985; Dame et al. 1986; Falgarone \& P\'{e}rault 1987; 
Solomon et al. 1987; Scoville et al. 1987; Leisawitz 1990; 
Fuller \& Myers 1992) and the clump mass distribution 
\nocite{Larson81,Leung+82,Myers83,Quiroga83}
\nocite{Sanders+85,Crovisier+85}
\nocite{Goldsmith+Arquilla85,Dame+86,Falgarone+Perault87}
\nocite{Solomon+87,Scoville+87,Leisawitz90,Fuller+Myers92}
(Myers, Linke \& Benson 1983; Blitz 1987; Carr 1987;
Loren 1989; Dickey \& Garwood 1989; Stutzki \& G\"{u}sten 1990; 
Lada, Bally \& Stark 1991; Nozawa et al. 1991; 
Langer, Wilson \& Anderson 1993; Williams \& Blitz 1993).  
\nocite{Myers+83,Blitz87,Carr87,Loren89}
\nocite{Dickey+Garwood89,Stutzki+Gusten90}
\nocite{Lada+91,Nozawa+91}
\nocite{Langer+93,Williams+Blitz93}

Theories to explain such laws have been presented for example by 
Larson (1981), Ferrini, Marchesoni \& Vulpiani (1983), 
Henriksen \& Turner (1984), Scalo (1987), Fleck (1988), 
Myers \& Goodman (1988), Mouschovias \& Psaltis (1995), 
Xie (1997), and Fleck (1996). Unfortunately, the results of these 
\nocite{Larson81,Ferrini+83}
\nocite{Henriksen+Turner84,Scalo87,Fleck88}
\nocite{Myers+Goodman88,Mouschovias+Psaltis95}
\nocite{Xie97,Fleck96}
investigations may be questioned because only the {\it projected} 
structure of molecular clouds is observed in the molecular emission line maps.
There is no way to de project the data since there in no relation between 
the observed radial velocity and the line--of--sight position of a feature.
Furthermore, these scaling laws suffer from serious confusion problems.
Operationally, clumps and other structures are identified by drawing closed
contours about local maxima in maps and/or in space--space--velocity data 
cubes. Many such features may consist of chance superpositions of 
physically unrelated structures.  Confusion is severe in crowded 
environments such as the interiors of giant molecular clouds and 
in the inner Galaxy (Scalo 1990; Maloney 1990; Issa, MacLaren \&
Wolfendale 1990; Adler \& Roberts 1992).
\nocite{Scalo90,Maloney90,Issa+90,Adler+Roberts92}

Since de--projection is not possible, the only way to correctly 
interpret observational data from molecular clouds is to project realistic 
theoretical models of their structure and dynamics into the observed 
data space.   Molecular clouds do contain magnetic fields
and supersonic motion.   Therefore, models of clouds must be 
based on solution of the equations that describe
magneto-hydrodynamic turbulence (e.g.\ Zweibel 1994). 
\nocite{Zweibel94}
We have produced numerical simulations of 
supersonic MHD turbulence that are solutions of the compressible 
3-D MHD equations in a regime of highly supersonic 
random motions (Padoan \& Nordlund 1997), and subsequently used a non-LTE Monte 
Carlo radiative-transfer code (Juvela 1997) to produce synthetic spectra 
\nocite{Juvela97}
of molecular transitions from our MHD data cubes. More than $5 \times 10^5$ spectra 
have been computed that may be compared with observational data. In this 
paper we describe the procedure used to build synthetic molecular clouds and present 
several results.  The MHD flow was computed in a 128$^3$ box with periodic
boundary conditions.   Radiative transfer calculations slightly degraded 
the spatial resolution to produce grids of 90$\times$90 spectra each 
containing 60 velocity channels.

\section{The numerical MHD simulations}

Padoan \& Nordlund (1997) have shown that a supersonic and
super-alfv\'{e}nic random flow is a good model for the dynamics and 
structure of molecular clouds and is consistent with the estimated magnetic
field strengths in cloud cores. In the present work, we use their numerical
run called experiment A (i.e., the super-Alfv\'{e}nic case). 
In this section we present the MHD equations and the numerical method 
used for their solution.

\subsection{The equations}

We solve the compressible MHD equations:

\def\vv{{\bf v}}
\def\jj{{\bf j}}
\def\bb{{\bf B}}
\def\lnr{\ln\rho}
\def\div{\nabla\cdot}

\begin{equation}
\label{0}
{\partial \ln\rho \over \partial t} + \vv \cdot \nabla\lnr = - \div \vv,
\end{equation}

  \begin{equation}
   {\partial{\vv} \over \partial t}
   + {\vv\cdot\nabla\vv}
  =
   - {P\over\rho} \nabla \ln P
   + {1\over\rho} {\jj} \times {\bb} 
   + {\bf f}_{\rm viscous}
   + {\bf f}_{\rm external},
  \label{1}
  \end{equation}

\begin{equation}
\label{4}
{\partial e \over \partial t} + {\vv} \cdot \nabla e = - {P \over \rho} \div 
{\vv} + Q_{\rm Joule} + Q_{\rm viscous} + Q_{\rm radiative},
\end{equation}

\begin{equation}
{\partial{\bb} \over \partial t} = 
\nabla\times (\vv\times\bb - \eta \nabla\times\bb)
\label{2}
\end{equation}

\begin{equation}
\jj = \nabla\times\bb,
\label{3}
\end{equation}

\noindent
plus numerical diffusion terms, and
with periodic boundary conditions. $\vv$ is the velocity, $\bb$ the
magnetic field, ${\bf f}_{\rm viscous}$ is the viscous force, 
${\bf f}_{\rm external}$ an external force ($=0$ in these particular
experiments), and $P = \rho T$ is the pressure at $T \approx$ const. 

We use an isothermal equation of state because the heat exchange is 
so efficient in molecular clouds, that the temperature remains low in most places.

\subsection{The code}

The code solves the compressible MHD equations on a 3D staggered mesh, 
with volume centered mass density and thermal energy, face centered 
velocity and magnetic field components, and edge centered electric currents 
and electric fields (Nordlund, Stein \& Galsgaard 1996).
\nocite{Nordlund+ea96para95}

The original code works with ``per-unit-volume'' variables; mass density,
momenta, and thermal energy per unit volume.  In the super-sonic 
regime relevant in the present application, we found it advantageous to 
rewrite the code in terms of ``per-unit-mass'' variables; $\ln\rho$, $u$, and 
$E=\frac{3}{2}\frac{P}{\rho}$.

We use spatial derivatives accurate to 6th order, interpolation accurate 
to 5th order, and Hyman's 3rd order time stepping method (Hyman 1979).
\nocite{Hyman1979}

Viscosity and resistivity are minimized using monotonic 3rd order 
hyper-diffusive fluxes instead of normal diffusive fluxes, and 
hydrodynamic and magneto-hydrodynamic shocks are captured by adding 
diffusivities proportional to the negative part of the velocity 
divergence, and resistivity proportional to the negative part of the 
cross-field (two-dimensional) velocity divergence.  Further details 
of the numerical methods are given by Nordlund, Galsgaard \&
Stein (1996) and Nordlund \& Galsgaard (1997).

\subsection{The Experiment}

Our synthetic molecular cloud data cubes are based on a numerical flow  
with $\beta_i=5.3$, where $\beta_i$ is the ratio between gas pressure and 
magnetic pressure at the initial time, $\beta_i=(P_{g}/P_{m})_i$. 
The initial density is assumed to be uniform.
The initial velocity field is random and purely solenoidal. 
The probability distribution is Gaussian and consists of only 
long wavelength modes.  In practice, this in done in Fourier space by 
including random  Fourier components with wave-numbers having a modulus 
$1\leq k\leq 2$.
The initial magnetic field is uniform, and is oriented parallel to 
the $z$ axis, ${\mathbf B}=B_0{\mathbf \hat{z}}$. No external forces 
are applied so the flow decays in time.  The magnetic energy 
is amplified by stretching of field lines and by compression immediately 
after the beginning of the simulation and the mean value of 
$\beta$ becomes $<\beta>\approx1.0$.  

The alfv\'{e}nic Mach number, ${{\cal{M}}_{A}}$ (), is initially 
${{\cal{M}}_{A}}\approx 10$, while the 
ordinary Mach number (i.e., the ratio of the 
flow rms velocity and the sound velocity) is initially ${\cal{M}} \approx 5$.

Details about the evolution of this flow, and the property of the
magnetic field (especially the relationship between volume density and 
field strength) may be found in Padoan \& Nordlund (1997).

\section{The Spectrum Calculations}

In order to make direct comparisons with observations we have
calculated J = 1--0 and J = 2--1 $^{12}$CO, $^{13}$CO and J~=~1--0 
CS spectra from model clouds where the
density and velocity fields are taken from the MHD calculations. Since
the LTE approximation might cause significant errors in the calculated spectra
we solve the non-LTE radiative transfer problem with a Monte Carlo
method (Bernes 1979). This way the effects of the velocity 
\nocite{Bernes79}
field and of the complex density structure are taken into account accurately.

\subsection{The Model Clouds}

The initial data cube consists of 128$^3$ cells for which the values of
density and macroscopic velocity are obtained from the MHD calculations.
These values were rescaled as described in section 4, in order to produce
the two cloud models, $A_{1}$ and $A_{2}$, summarized in Table~1. The cloud model 
$A_{1}$
has average gas density of 400\,cm$^{-3}$ and rms velocity 1.7\,km\,s$^{-1}$;
model $A_{2}$ density of 100\,cm$^{-3}$ and rms velocity 3.4\,km\,s$^{-1}$. The 
largest density values are around 1.0$\cdot$10$^5$\,cm$^{-3}$ in both models,
and  the kinetic temperature is set to 10\,K.

In order to speed up the radiative transfer calculations the data was
re sampled into smaller data cubes consisting of 90$^3$ cells. New density
and velocity values were calculated by linear interpolation of the
larger data set. The velocity dispersion between neighboring cells in the
original data cube was calculated and used to approximate the
turbulent line widths within each cells of the new data cubes. After 
the thermal line broadening corresponding to the kinetic temperature of 
10\,K in hydrogen was added to that line width,  the mean intrinsic line width 
was found to be $\sim$0.7\,km\,s$^{-1}$. The kinetic temperature and 
the molecular abundances were kept constant throughout the clouds. 
The abundances relative to the abundance of H$_2$
were 5.0$\cdot 10^{-5}$ for CO, 1.0$\cdot 10^{-6}$ for $^{13}$CO and
5.0$\cdot 10^{-9}$ for CS.

\subsection{The radiative transfer calculations}

The radiative transfer calculations were carried out with a Monte Carlo
program that is well suited for this kind of a study where the cloud
contains complex density and velocity fields.  With this method, 
variations in kinetic temperature, molecular abundances, and external 
radiation fields may be traced explicitly. However, in the present
calculations, the temperature and relative abundances are kept 
constant and no external radiation field is considered, apart from 
the 2.73~K background radiation. 

Our program is a generalization of the one-dimensional Monte Carlo method
(Bernes 1979) to three dimensions. The model cloud is divided into cells
in which physical properties of the gas are assumed to be constant and in
our case the cells are small cubes. There are, however, important
differences between our program and the normal Monte Carlo method and it 
is therefore necessary to explain some principles of the implementation. 
A detailed description is given elsewhere (Juvela 1997).

In the normal Monte Carlo method the radiation field is simulated with
photon packages representing a number of real photons. The packages are sent
from random locations in the cloud towards random directions. Each package
is followed out from the cloud and interactions between the molecules and
the photons contained in the package are counted. When sufficient number of
model photons has been simulated this information is used to solve in each
cell the number of molecules on different energy levels and the iteration
continues with the new level populations.

We have used a modified method in which the radiative transfer is simulated
along lines crossing the cloud uniformly.
Photon packages containing cosmic microwave background photons are 
sent into the cloud along these lines.  Photons emitted by a cell are added
to the incident photons. The photon numbers, in particular the number
of photons absorbed within the emitting cell, are calculated explicitly. 
This is important if some of the cells are optically
thick. In that case the normal Monte Carlo method requires generation of a large
number of emission events within each of the optically thick cells.

The radiation field was simulated using photon packets containing complete
intensity distributions of the simulated lines. The lines were divided into
60 velocity channels that were treated separately in the calculations and
for this reason there are no random errors associated with the generation of
model photons with random Doppler shifts. The total velocity range included
in the calculations was 10\,km\,s$^{-1}$ for the model $A_{1}$ and
20\,km\,s$^{-1}$ for the model $A_{2}$ and the channel widths were
correspondingly 0.17\,km\,s$^{-1}$ and 0.33\,km\,s$^{-1}$. The channel
widths are much smaller than the line widths and less or equal to the
smallest intrinsic line widths found in the cells. The chosen velocity
discretization is therefore not expected to affect the results of the
calculations.

The number of model photons was about 200\,000 per iteration.  Each
model photon contains the intensity distributions of all transitions.
Each model photon also represents the emission from all points
along a line of sight through the cloud (instead of just from one point). 
Therefore, the number of model photons can be much less than 
required by the normal Monte Carlo method and in this 
case 200\,000 model photons are found to give sufficiently accurate results.
New level populations are solved in each iteration from the equilibrium
equations and convergence was checked by computing the relative changes
in the populations of the six lowest energy levels. The iterations 
stopped when this change, averaged over all cells in the model cloud, was
$\sim$1.0$\cdot$10$^{-3}$. In some cells the convergence was much worse than
the average. These cells, however, have low densities and are
therefore not expected to contribute to the spectra. Furthermore, 
the calculated spectra depend only on populations on the first three 
excitation levels and these could be more reliably determined 
than the number of molecules on the sparsely populated upper levels.

The collision coefficients for CO and $^{13}$CO were taken from Green \&
Thaddeus (1976) and Green and Chapman (1978). For these molecules six 
\nocite{Green+Thaddeus76,Green+Chapman78}
energy levels were included in the calculations, sufficient for 
excitation temperatures less than 10\,K. The CS collision
coefficients are from Turner et al. (1992). For CS, the lowest 
\nocite{Turner+92}
eight energy levels were included to ease the memory requirements.  
With T$_{\rm ex}$=10\,K the number of molecules in the level
$J$=7 is less than one percent of the number of molecules in the
$J$=1 or $J$=2 states. Since levels are sub-thermally populated, 
the omission of upper energy levels will not affect the results
of our calculations.

\subsection{The Spectra}

The level populations solved with Monte Carlo simulation were used to 
calculate molecular line spectra along different direction through the cloud. 
Spectra were computed on a 90$\times$90 grid. 
Three separate grids of spectra were calculated to correspond to lines
of sight perpendicular to the faces of the MHD data cube.
Four other grids were calculated along the diagonal directions. 
In these cases the maps of 90$\times$90 spectra do not extend over the 
whole projected cloud area. Each spectrum corresponds to the emitted intensity 
calculated along a particular line of sight through the cloud.  These spectra
have not been convolved to synthesize a realistic beam.  The spectra contain
60 velocity channels as in the Monte Carlo simulation. 

For $^{13}$CO and CS the optical depths of individual cells were
small ($\tau \ll 1$).  The average optical depth in $^{12}$CO, however, 
was about 0.4 per cell and many cells were optically thick.   We have tested
the effects of degrading the spacing in the final grids from 90$^3$ to 
45$^3$ cells.  Despite the high optical depth of some cells,
this procedure was found to have only a negligible effect on
the calculated line intensities.  
Only small differences could be seen between these two models. 
The maximum differences in individual velocity channels were 
only a few percent.  Therefore, the difference between discretization into 
90$^3$ or 128$^3$ cells is unimportant for the spectrum calculations.

The results were compared with spectra calculated assuming LTE conditions.
The comparison showed that in this case the LTE assumption would be 
unsuitable.

\section{Synthetic Molecular Clouds}

We have computed 70 grids of 90$\times$90 spectra in the 
J=1$\rightarrow$0 and J=2$\rightarrow$1 lines of $^{12}$CO, 
the J=1$\rightarrow$0 and J=2$\rightarrow$1 $^{13}$CO,  and 
the J=1$\rightarrow$0 CS.  Model $A_{1}$ describes 5~pc clouds
while model $A_{2}$ describes 20~pc clouds and Table~1 summarizes  
their main parameters.  For each model we calculate maps corresponding 
to 7 different lines--of--sight through the 3 projected faces and 
four projected diagonals of the model data cube.  

All clouds have the same mean column density, 
N(H$_{2}$)=6$\times$10$^{21}$ cm$^{-2}$, a typical value 
for observed clouds.  Models $A_{1}$ and $A_{2}$ have 
mean densities $\langle n\rangle=400$~cm$^{-3}$ and $\langle n\rangle=100$~cm$^{-3}$ 
respectively. The three-dimensional rms velocity 
(not the line width), $\sigma_v=1.7$ \kms\ 
in models $A_{1}$ and 3.4 \kms\ in models $A_{2}$. These rms velocities 
produce line widths in agreement with observations. The numerical 
flow has an rms Mach number $\le$5.  It is easy to re scale the three 
dimensional velocity fields to other values of the rms Mach  number. 
The density field of a flow with a larger Mach number may also
be obtained from statistical properties of random supersonic 
(and super-Alfv\'{e}nic) flows found by Nordlund \& Padoan (1997), and 
\nocite{Nordlund+Padoan97sst}
proven to be in agreement with stellar extinction determinations 
(Lada et al. 1994) in Padoan, Jones \& Nordlund (1997a), and 
\nocite{Lada+94}
Padoan \& Nordlund (1997). The density field is well approximated 
by a {\it log--normal distribution}
whose standard deviation is 0.5 times the rms Mach number of the flow. 
Therefore, if the mean temperature and rms velocity in molecular clouds are 
known, the statistics of the density field are completely determined. 
To make the density field consistent with the rescaled velocity field, we
scale the logarithm of the density to the correct standard deviation and 
mean, and then take the exponential.  The result is a 
rescaled density field with the same mean as the original one 
but with a larger standard deviation, consistent with the velocity field.

\section{Results}

\subsection{Integrated Temperature Maps and Distributions}

The complex three-dimensional density field in the 
numerical flow is illustrated in Fig.~1, using
volume projections with different opacity levels.  
The morphology appears `filamentary' and the density contrast is 
more than five orders of magnitude. The distribution is Log--Normal 
as in previous experiments (Nordlund \& Padoan 1997; Padoan, Jones \& Nordlund 1997). 
\nocite{Nordlund+Padoan97sst,Padoan+97ext}
Fig.~\ref{f2} shows the cumulative volume and mass fractions as 
functions of density.  This allows the present model of the density 
field to be related to previous clumpy models of molecular clouds used 
to produce synthetic spectra. For example, in the 5~pc model,
80\% of the volume has a density below the mean ($n<$400 cm$^{-3}$), 
and therefore will not give a significant contribution to CO 
or CS emission.  Nevertheless, 30\% of the total mass is located in 
only 1.4\% of the volume and is in regions ten times 
denser than the mean density ($n>$4000 cm$^{-3}$). Thus, a 
significant fraction of the mass but a very small fraction of the volume will 
contribute to the CO emission. The density distribution in the 20~pc 
model is even more intermittent because of the larger
rms Mach number of the flow.  Almost 90\% of its volume is below the average
density ($n<$100 cm$^{-3}$) and about 20\% of the mass has 
$n>$4000~cm$^{-3}$.  These dense regions occupy only 0.2\% of the 
total volume.  It is difficult to compare this complex 
and intermittent density field to previous models built as 
ensemble of clumps because in the present case there is a continuous 
distribution of density values from $n=$0.6 cm$^{-3}$ to 
$n=$10$^5$ cm$^{-3}$ in the 5~pc model and 
from $n=$0.01 cm$^{-3}$ to $n=$10$^5$ cm$^{-3}$ in the 20~pc model.
Figs.~3-5 show several examples of integrated 
intensity maps of J=1$\rightarrow$0 transition of 
$^{12}$CO, $^{13}$CO and CS.  It is apparent that the 
CS emission has a lower surface filling factor than the 
$^{13}$CO emission and the $^{13}$CO emission has a lower 
surface filling than the $^{12}$CO emission.

The integrated temperature maps are reminiscent of observed clouds.
Filaments and clumps are apparent as well as features that look like 
bubbles and which might be interpreted as shells due to the star
formation activity.   In our models, this morphology 
can only be due to the complex system of shocks 
that develops in the random supersonic flow. It is remarkable that 
an extremely rich structure is produced even without the effect of 
gravity, stellar radiation, or stellar outflows. 
The density contrast is large and comparable to that estimated in dark clouds
and the velocity dispersions are realistic.

The probability distribution of the integrated intensity is plotted in 
Fig.~\ref{6} for the j=1$\rightarrow$0 transitions of all three 
molecules in the 5~pc models.  The distributions are obtained 
from 7 synthetic clouds.  While $^{12}$CO has a rather symmetric 
(nearly Gaussian) distribution,  $^{13}$CO and CS are characterized by 
long tails because they sample the column density 
which is by itself intermittent.

\subsection{Spectral Line Profiles}

The study of the spectral line profiles is interesting because of previous
attempts to use their shape to probe the presence of turbulence in dark
clouds. Falgarone \& Phillips (1990) studied molecular line profiles
\nocite{Falgarone+Phillips90}
of different sources, and found excess emission in the line wings,
relative to a Gaussian distribution (see also Blitz, Magnani \& Wandel 1988).
\nocite{Blitz+88}
Miesch \& Scalo (1995) looked at the histograms of emission line centroids,
\nocite{Miesch+Scalo95}
and found nearly exponential tails in many cases. Lis et al. (1996) 
\nocite{Lis+96}
studied the statistics of line centroids and of centroid increments, 
and found non-Gaussian behavior especially in the histograms of the 
velocity centroid increments. There is therefore observational evidence of 
intermittency in the statistics of velocity fields in molecular clouds.

In Fig.~\ref{7} we plot the average spectra of the transitions 
J=1$\rightarrow$0 and J=2$\rightarrow$1, of $^{12}$CO and $^{13}$CO. 
The average is made over seven synthetic maps of the same cloud model, 
for 5~pc (left), and 20~pc (right), that is to say over 56700 individual 
spectra. While the $^{13}$CO spectra are very close to Gaussian, 
the $^{12}$CO spectra show saturation and a little bit of 
self-absorption. All spectra are very smooth and centrally peaked.

Individual spectra are shown in Figs.~\ref{10}-\ref{12}. We have plotted 
only subsets of 20$\times$20 spectra, while the maps contain
90$\times$90 spectra, in order to better visualize the profiles.  
Individual line profiles often shows multiple velocity components,
and this is especially true for $^{12}$CO. The $^{12}$CO profiles in high
intensity spots are clearly saturated, and this occurs at a radiation 
temperature of about 7 K. It is difficult to describe the shapes of 
individual line profiles, because while some are nearly Gaussian, with 
an excess wing emission (as in the observational analysis by 
Falgarone \& Phillips), some have double or multiple components. 

In order to quantitatively illustrate the shape of the individual
spectral profiles, we have plotted in Figs.~\ref{13}-\ref{14} the 
distributions of their first four statistical moments: mean velocity, 
velocity dispersion, skewness, and kurtosis. The vertical dashed lines 
in the plots mark the value of the correspondent statistical moment 
in the averaged spectrum (Fig.~\ref{7}).  The saturation in the 
$^{12}$CO spectra manifests itself in lower values of
kurtosis relative to the $^{13}$CO spectra. 
Both skewness and kurtosis in the 20~pc models (not shown) span a range of 
values twice as large as in the 5~pc models, although those 
statistical moments are normalized with the 
velocity dispersions. The broad distribution of the values of kurtosis around
the Gaussian value of 3, is due to the fact that while some lines are just 
Gaussian plus excess wing emission (kurtosis larger than 3), some have 
multiple components that tend to flatten the profile (kurtosis less than 3).

\subsection{Line-Width Versus Integrated Temperature}

Fig.~\ref{17} shows the relation between integrated 
temperature and equivalent width, defined as the integrated temperature 
divided by the maximum temperature, for $^{13}$CO, J=1$\rightarrow$0. 
Heyer, Carpenter \& Ladd (1996) have drawn attention to this relation, 
\nocite{Heyer+96}
because it could be used as a way to probe the importance of magnetic 
fields for the dynamics of molecular clouds. They claim that Alfv\'{e}n 
wave models of molecular cloud motions would generate a decreasing 
line-width with increasing column density.
We show here that the numerical super-Alfv\'{e}nic flow does produce an 
increasing line-width with increasing integrated temperature,
as found in the observations. 
Plots are obtained as an average on several maps. Plots for 
single maps are qualitatively the same.

\subsection{Line Intensity and Line Width Ratio}

The observed line intensity ratios of different CO transitions
are found to be approximately constant.
Falgarone \& Phillips (1996) find $R_{CO}(2-1/1-0)=0.62\pm0.08$,
\nocite{Falgarone+Phillips96}
constant in space, and also across line profiles, in a
portion of a cloud edge in the Perseus-Auriga complex.
In Fig.~\ref{18} the average $^{12}$CO J=1$\rightarrow$0
(continuous line) and J=2$\rightarrow$1 (dotted line)
spectra are plotted for both the 5 pc and the 20 pc models.
The J=2$\rightarrow$1 temperature is divided by
0.62, that is the line ratio found by Falgarone \& Phillips
(1996). Fig.~\ref{18} shows that the synthetic spectra
from our model clouds have approximately the same intensity
ratio has the observed ones. The same is true for the ratios
of velocity integrated temperature of single lines--of--sight, 
shown in Fig.~\ref{19}. The scatter in the plot
is comparable to the observed scatter, although our models
produce a ratio that slightly grows with increasing 
$^{12}$CO J=1$\rightarrow$0 integrated temperature.

Finally, the ratios of $^{12}$CO J=2$\rightarrow$1 to
$^{13}$CO J=2$\rightarrow$1 line--widths are plotted
in Fig.~\ref{20}, where the continuous horizontal line
is again the value found by \citetext{Falgarone+91}.

\section{Discussion}

In this work we have illustrated the procedure of 
obtaining synthetic molecular spectra that may be used 
to interpret observational data on molecular clouds. The spectrum
calculations are based on cloud models that are consistent 
with the physical conditions in molecular clouds. We have 
not included gravity in the cloud models, because one 
of our aims is to show that magneto--hydrodynamics itself, 
without gravity, can successfully reproduce the structure and 
kinematics observed in molecular clouds.  This is due to 
the fact that the velocity field is highly supersonic and turbulent.
Under these conditions the flow develops a 
complex system of shocks with very high density contrast, because 
the cooling time in the molecular gas is so short. Therefore
molecular clouds are fragmented primarily by a random system of shocks, and not by 
gravitational instability, although the formation of protostars inside 
dense regions of shocked gas is eventually a gravitational 
collapse. The statistical properties that arise from this scenario 
of {\it turbulent fragmentation} were 
shown to be in agreement with the statistics of the density field in 
the dark cloud IC 5146 (Padoan, Jones \& Nordlund 1997),
\nocite{Padoan+97ext}
studied by Lada et al. (1994) with near-infrared imaging. 
\nocite{Lada+94}
Padoan \& Nordlund (1997) have shown that the same scenario is also 
consistent with the estimates of magnetic field--strength in molecular clouds. 
A direct result of this picture is also the 
successful derivation of the stellar initial mass function 
(Padoan, Nordlund \& Jones 1997b).
\nocite{Padoan+97imf}

There is remarkable qualitative agreement between both the synthetic
spectra and the images generated by integrating the spectra over narrow
velocity ranges.  For example, the CO maps of the Orion region
presented by Bally et al. (1987) and by Bally, Langer \& Liu (1991) 
\nocite{Bally+87,Bally+91}
exhibit filamentary morphology and contain voids of various sizes 
that closely resemble the images presented in this paper.   
A comparison between the theoretical data cubes and 
$^{13}$CO observations of the Perseus molecular cloud obtained 
with linear resolution comparable to the model calculations 
shows close agreement in a variety of statistical measures 
(Padoan et al. 1997). 

A detailed comparison of the theory with the observations is now 
possible and will be presented in future works. 
The results that have been shown here are certainly
reminiscent of the observational data. Some of them are:

\begin{itemize}
 
\item The complex filamentary and clumpy structure of the 
integrated intensity maps.

\item The long tail (intermittency) in the probability distribution 
of the integrated
intensity of $^{13}$CO and CS.

\item The smooth centrally peaked CO average spectra.

\item The presence of multiple components and intermittent wings in 
spectra from individual lines of sight.

\item The distributions of statistical moments from individual spectra.

\item The relation between integrated intensity and line--width.

\item The line intensity and line--width ratios.

\end{itemize}

All these properties of molecular clouds are nicely reproduced by the numerical
flows, even if their initial conditions (eg uniform density and 
magnetic fields, solenoidal large scale velocity field) are not 
necessarily a good description of molecular clouds in any phase of their evolution. 
The success of the numerical models is due to the fact that the 
system almost immediately loses its memory of the initial conditions.
In chaotic dissipative systems such as the present one, the position and 
velocity of a particle at any given time are, on the one hand,
extremely sensitive to the initial conditions, and are therefore
individually unpredictable (initial errors
grow exponentially with time). On the other hand, such systems
also develop some order, or self--organization. The self--organization is
apparent in a statistical description (e.g. scaling laws and probability
distribution functions). Thus, while
a turbulent flow is extremely sensitive to the initial conditions
from a mechanic point of view, the initial conditions are almost irrelevant
for its statistical description. This is the reason why the numerical
modeling is able to predict statistical properties of turbulent
flows that are universal, and also measured in molecular clouds. Ironically,
the success of the numerical modeling is due to the very large Reynolds
number of the motions in molecular clouds, which is usually believed to be the
most serious limitation of numerical simulations (eg Scalo 1987).
\nocite{Scalo87}

\acknowledgements

This work has been partially supported by the Danish National Research 
Foundation through its establishment of the Theoretical Astrophysics Center.
Computing resources were provided by the Danish National Science 
Research Council,
and by the French `Centre National de Calcul Parall\`{e}le en 
Science de la Terre'.
PP is grateful to the Center for Astrophysics and Space 
Astronomy (CASA), in
Boulder (Colorado), for the warm hospitality offered during the period in which
this paper has been written.
JB acknowledges support from NASA grants NAGW-4590 (Origins) and 
NAGW-3192 (LTSA).
The work of MJ was supported by the Academy of Finland Grant No. 1011055.



\clearpage

{\bf Figure and Table captions:} \\

{\bf Table 1:}  Model physical parameters. \\

{\bf Figure 1:} 3--D density field from a numerical 
simulation. Upper panel: Projection with high opacity to show the 
external faces. Lower panel: Projection with low opacity to 
show the internal structure.  \\

\begin{figure}
\centering
\leavevmode
\epsfxsize=0.5
\columnwidth
\epsfbox{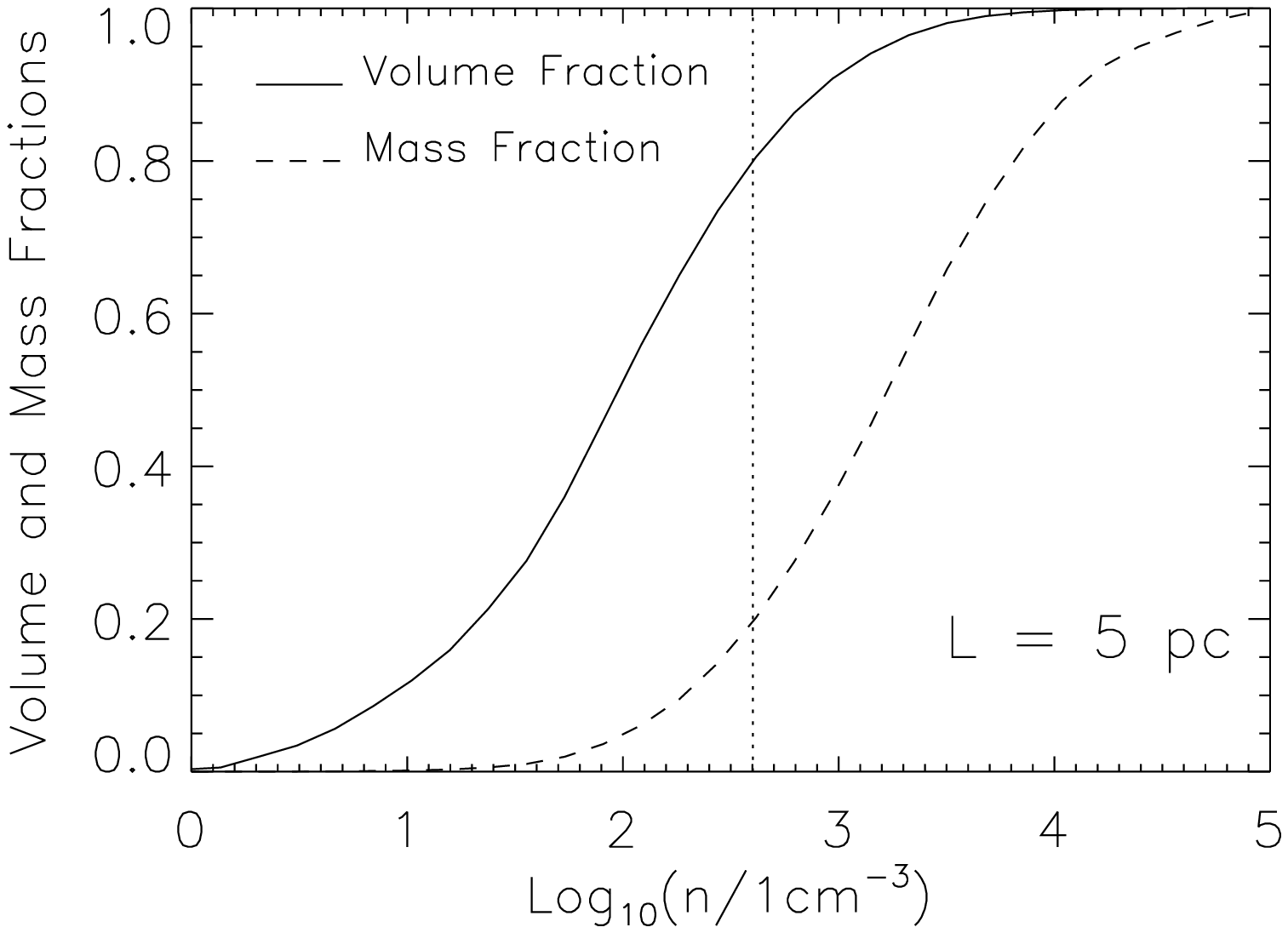}
\epsfxsize=0.5
\columnwidth
\epsfbox{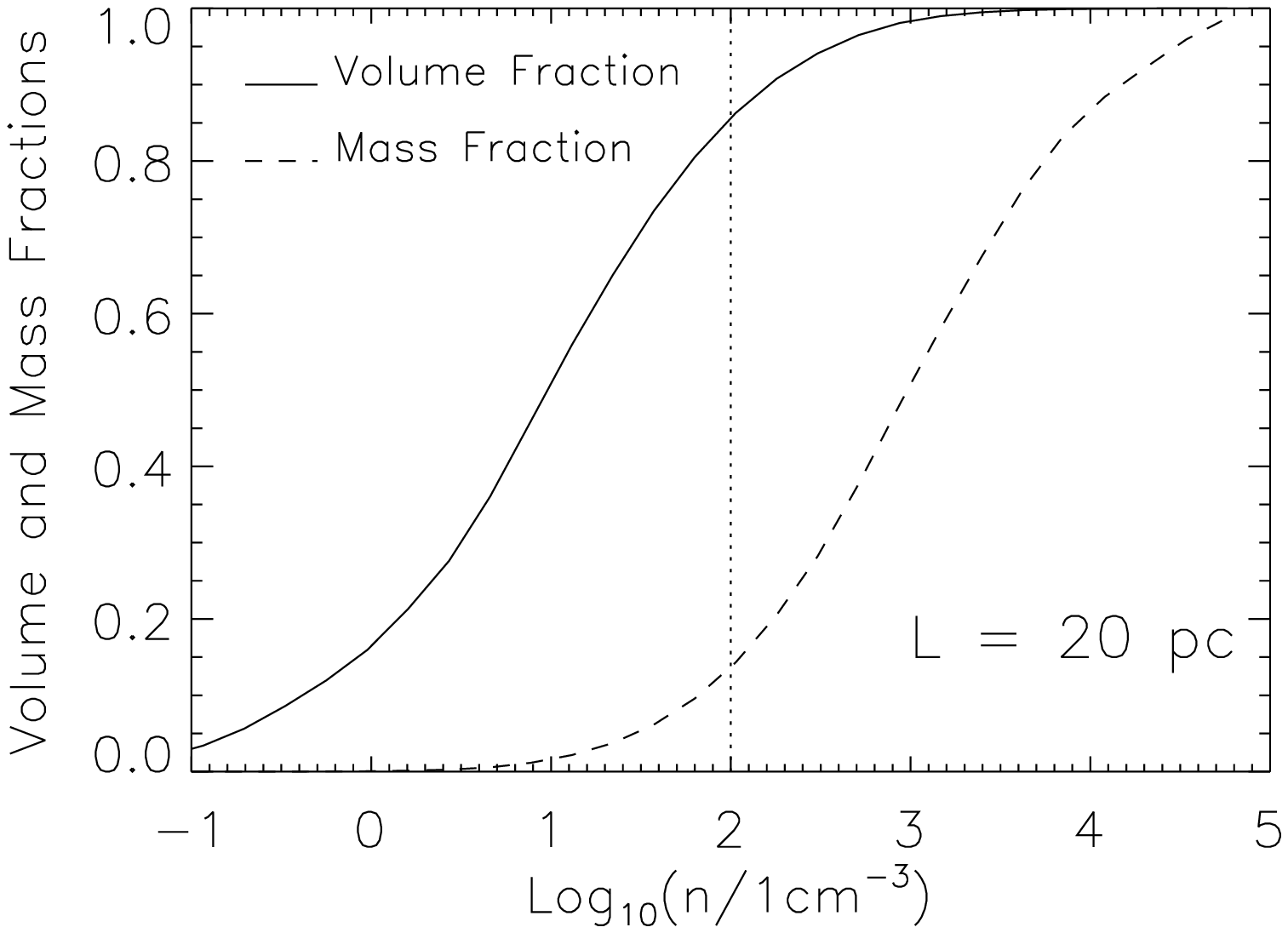}
\caption[]{}
\label{f2}
\end{figure}

{\bf Figure \ref{f2}:}  Volume and Mass Fractions in the 5pc (left) 
and 20pc (right) 3-D density distribution, used to produce the 
synthetic clouds. The vertical dotted line marks the mean density. \\

{\bf Figure 3:} Synthetic integrated temperature maps of $^{12}$CO, 
J=1$\rightarrow$0, for the 5pc clouds. Each map is obtained with 
90$\times$90 spectra. The intensity scale of the map is proportional 
to the integrated temperature. \\

{\bf Figure 4:}  As in Fig.~\ref{3}, but for $^{13}$CO, 
J=1$\rightarrow$0.\\

{\bf Figure 5:}  As in Fig.~\ref{3}, but for CS, 
J=1$\rightarrow$0.\\

\begin{figure}
\centering
\leavevmode
\epsfxsize=1.0
\columnwidth
\epsfbox{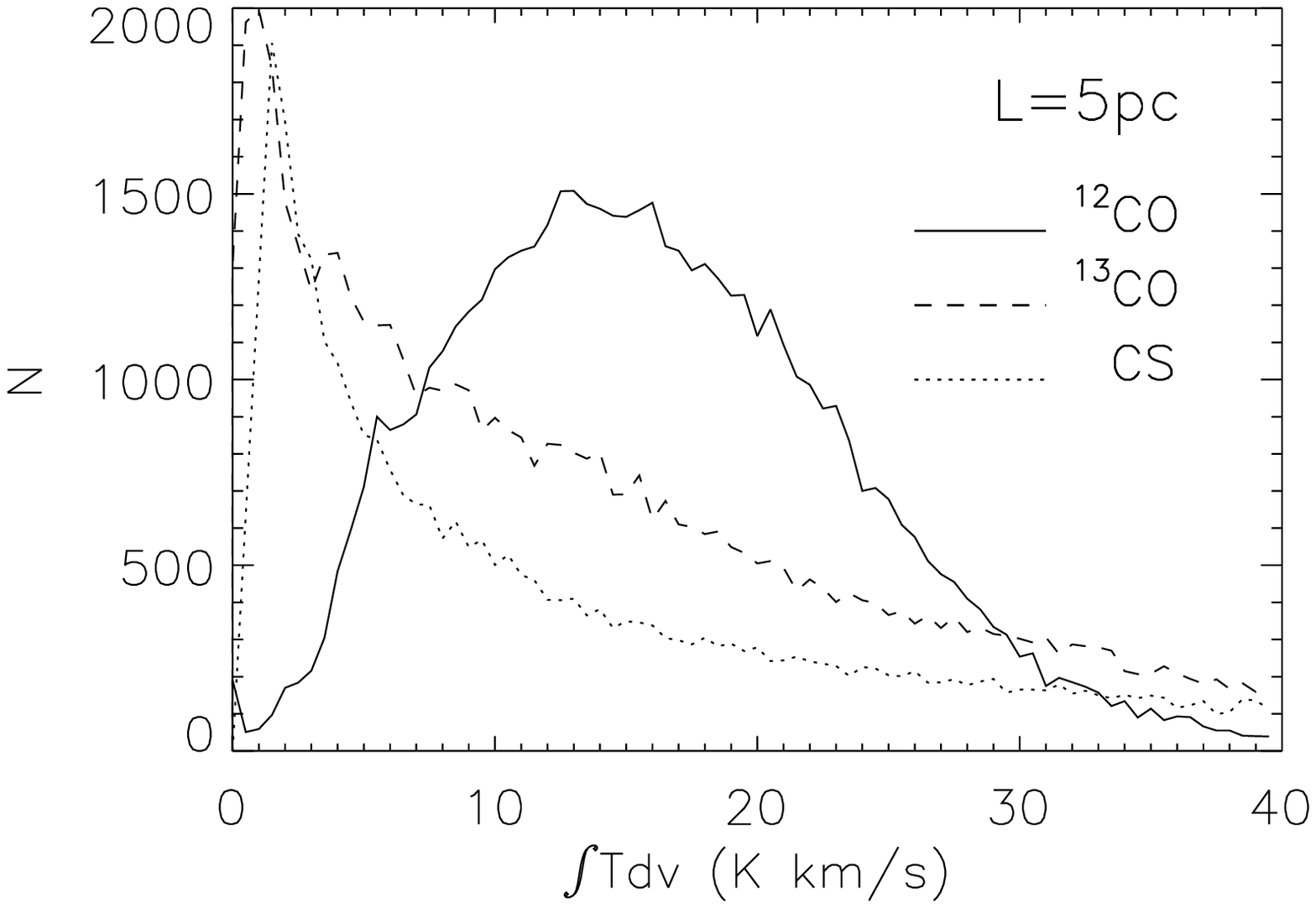}
\vspace{5cm}
\caption[]{}
\label{6}
\end{figure}

{\bf Figure \ref{6}:}  Probability distribution of the integrated 
temperatures. The $^{13}$CO temperature has been multiplied by a factor 
5.0, and the CS temperature by a factor 100.0.\\

\begin{figure}
\centering
\leavevmode
\epsfxsize=0.5
\columnwidth
\epsfbox{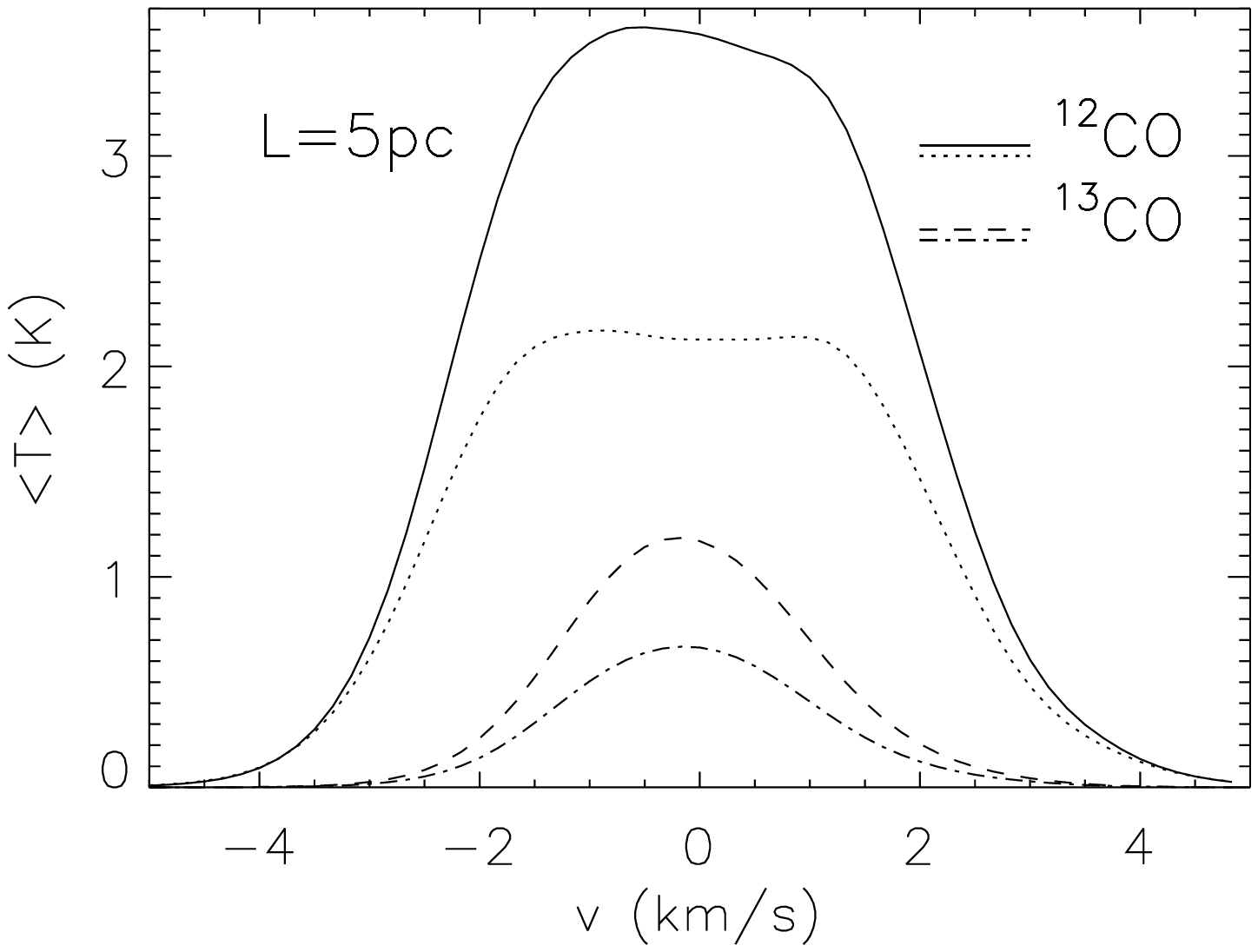}
\epsfxsize=0.5
\columnwidth
\epsfbox{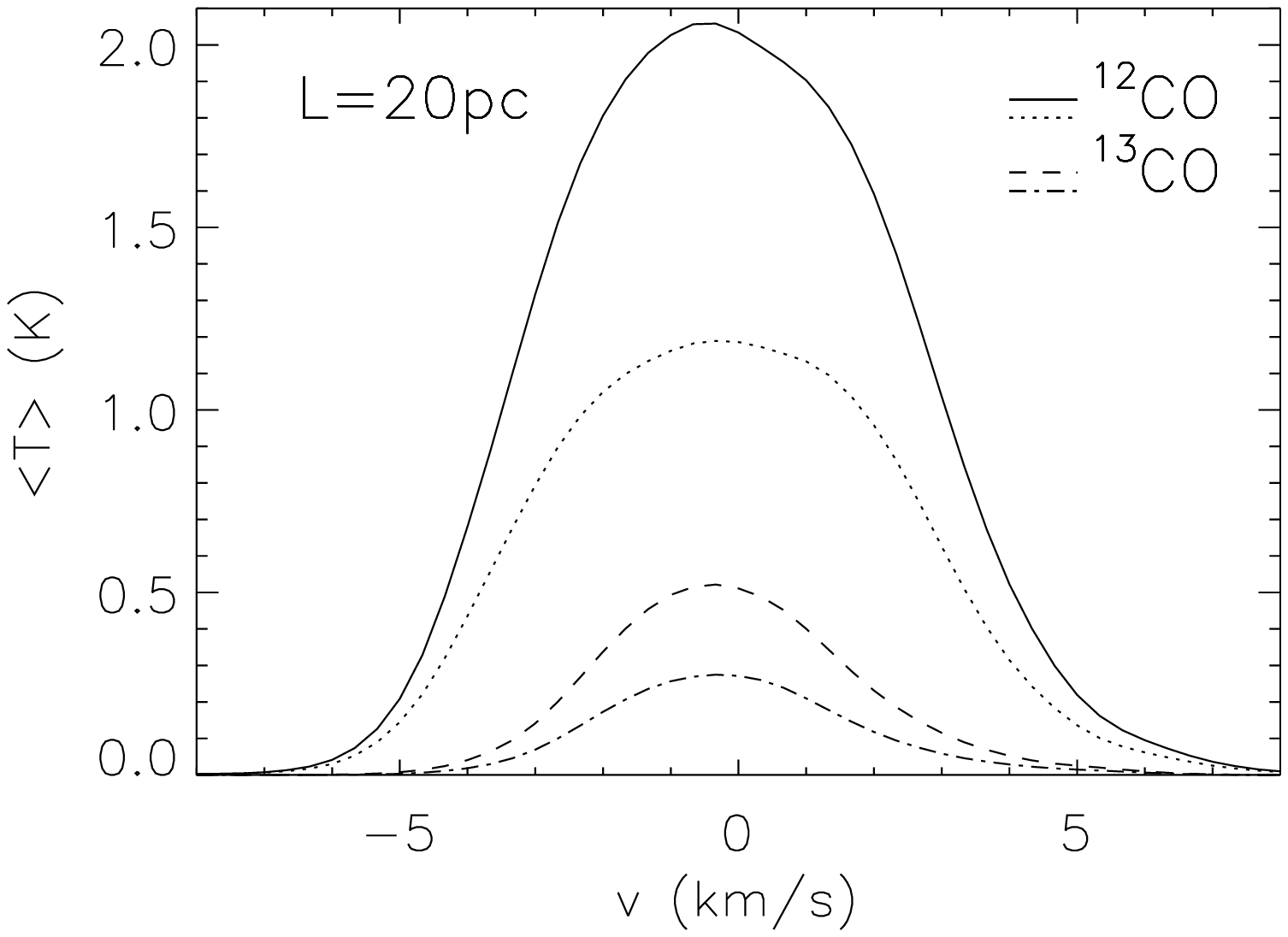}
\caption[]{}
\label{7}
\end{figure}

{\bf Figure \ref{7}:}  Average spectra for the 5pc (left) and 
20pc (right) cloud models. From top to bottom:
$^{12}$CO, J=1$\rightarrow$0, $^{12}$CO, J=2$\rightarrow$1, 
$^{13}$CO, J=1$\rightarrow$0, and $^{13}$CO, J=2$\rightarrow$1. 
Each of the four spectra is obtained averaging the spectra of 
seven synthetic clouds, that is 56700 spectra, with 60 velocity 
channels.\\



\begin{figure}
\centering
\leavevmode
\epsfxsize=1.0
\columnwidth
\epsfbox{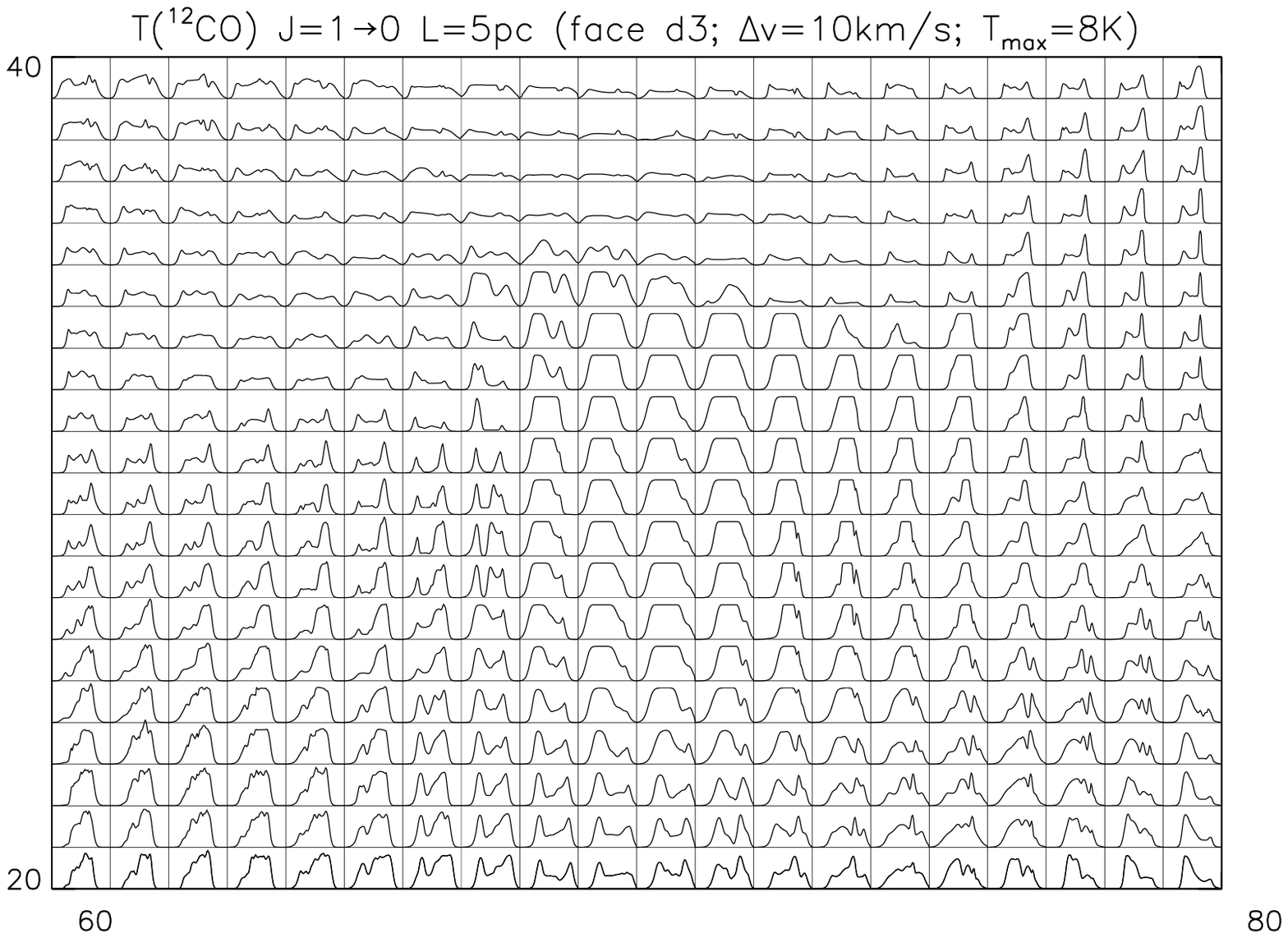}
\caption[]{}
\label{10}
\end{figure}

{\bf Figure \ref{10}:}  20$\times$20 spectra subset of $^{12}$CO, 
J=1$\rightarrow$0. A high intensity region of a 5pc model has 
been selected. For each spectrum only a 10 km/s
velocity range has been plotted, and the maximum 
temperature is 8 K. \\

\begin{figure}
\centering
\leavevmode
\epsfxsize=1.0
\columnwidth
\epsfbox{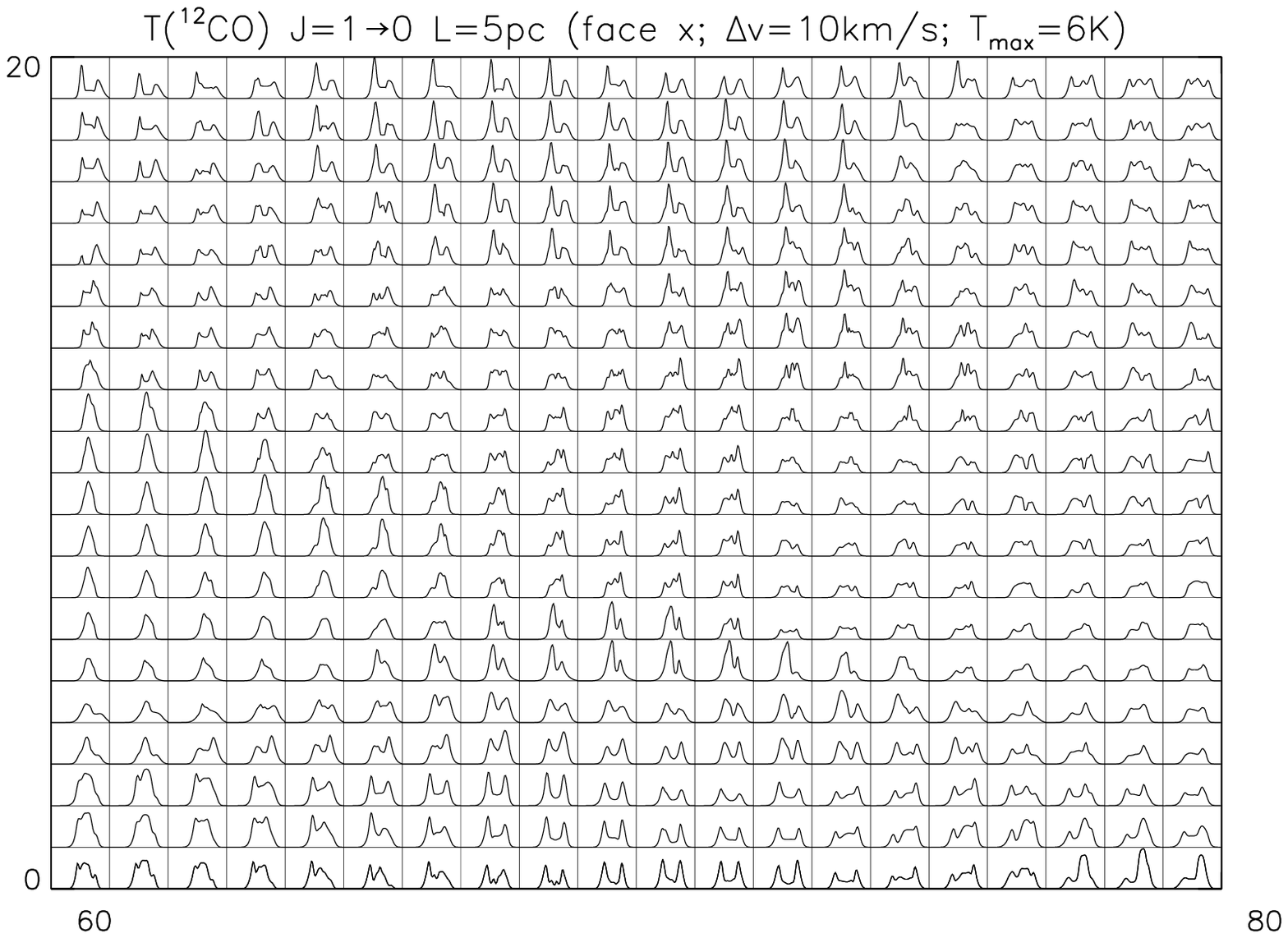}
\caption[]{}
\label{11}
\end{figure}

{\bf Figure \ref{11}:} 20$\times$20 spectra subset of $^{13}$CO, 
J=1$\rightarrow$0. A low intensity region of a 5pc model has 
been selected. For each spectrum only a 10 km/s
velocity range has been plotted, and the maximum 
temperature is 6 K. \\

\begin{figure}
\centering
\leavevmode
\epsfxsize=1.0
\columnwidth
\epsfbox{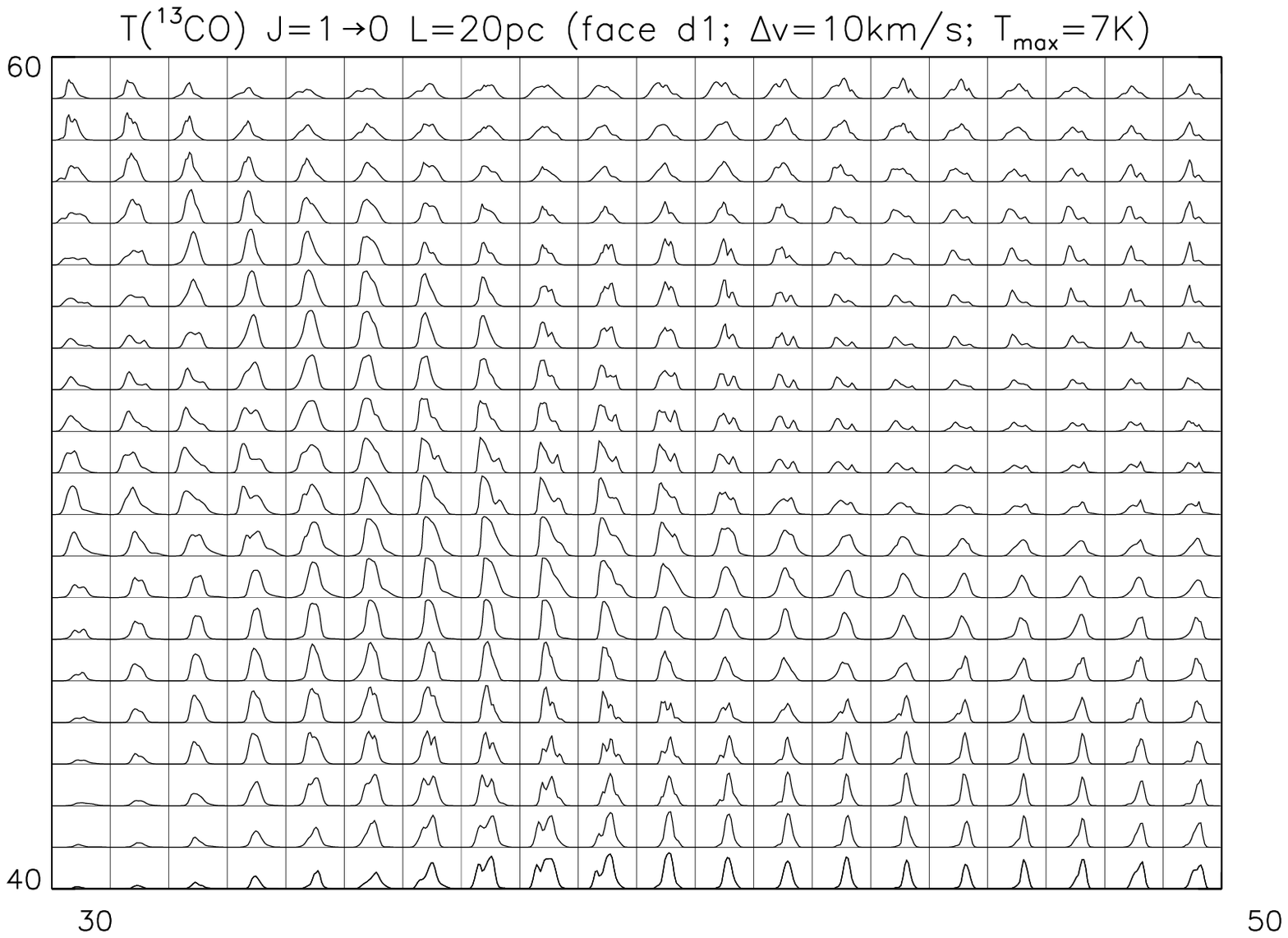}
\caption[]{}
\label{12}
\end{figure}

{\bf Figure \ref{12}:} 20$\times$20 spectra subset of $^{13}$CO, 
J=1$\rightarrow$0. A high intensity region of a 20pc model has 
been selected. For each spectrum only a 10 km/s
velocity range has been plotted, and the maximum temperature 
is 7 K. \\

\begin{figure}
\centering
\leavevmode
\epsfxsize=1.0
\columnwidth
\epsfbox{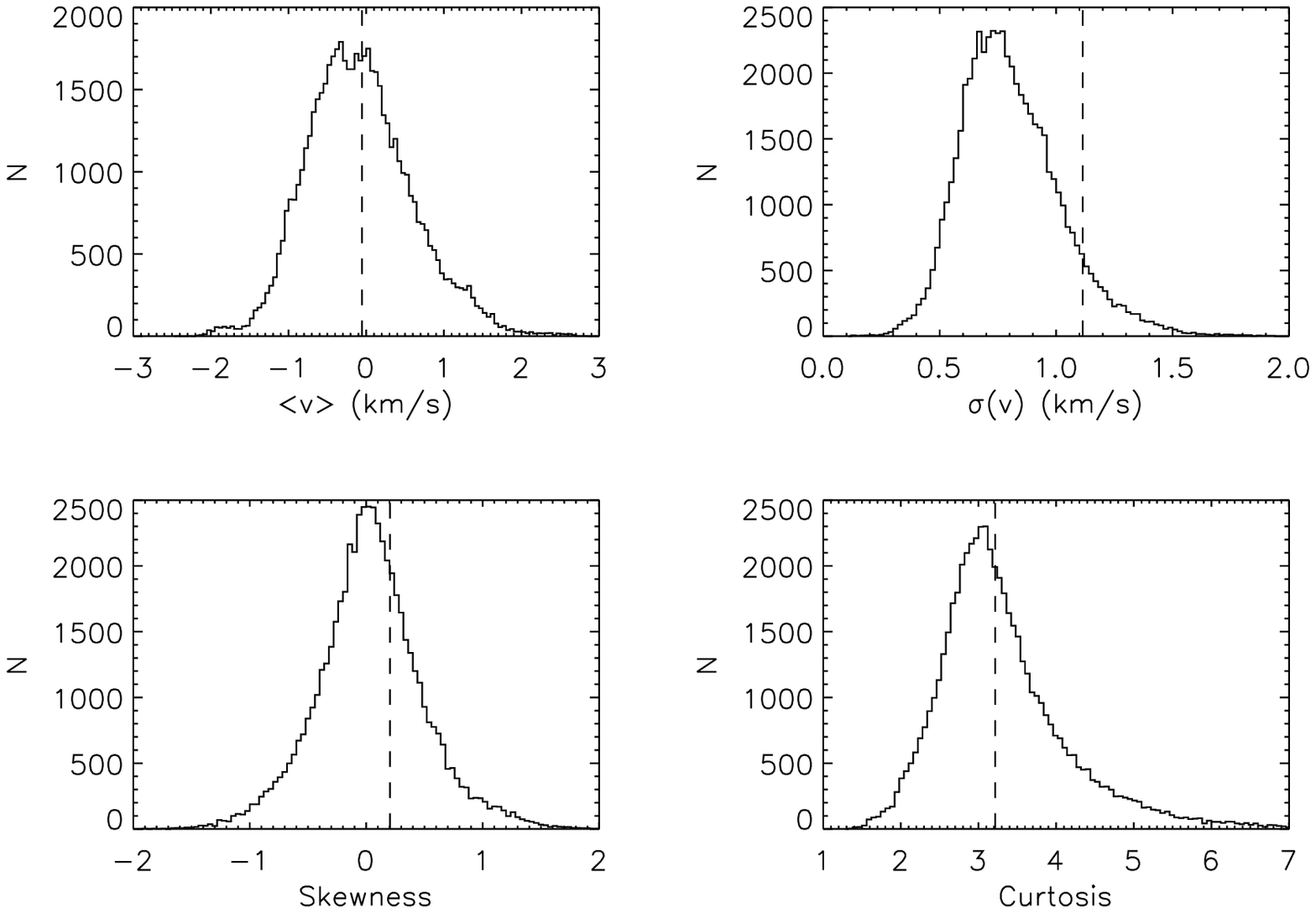}
\caption[]{}
\label{13}
\end{figure}

{\bf Figure \ref{13}:}  Distributions of statistical moments 
(centroid velocity, velocity dispersion, 
skewness, and kurtosis), for all the $^{13}$CO, 
J=1$\rightarrow$0, spectra of 7 5pc 
cloud models (56700 spectra). The clouds are 
the same used to produce the mean spectra 
plotted in Fig.~\ref{7}. The values of the same four 
statistical moments for the mean 
spectrum is plotted as vertical dashed lines. 
While the mean spectrum is very close to 
Gaussian, the individual spectra can have rather 
complex shapes, leading to values
of the moments quite far from the Gaussian ones.\\

\begin{figure}
\centering
\leavevmode
\epsfxsize=1.0
\columnwidth
\epsfbox{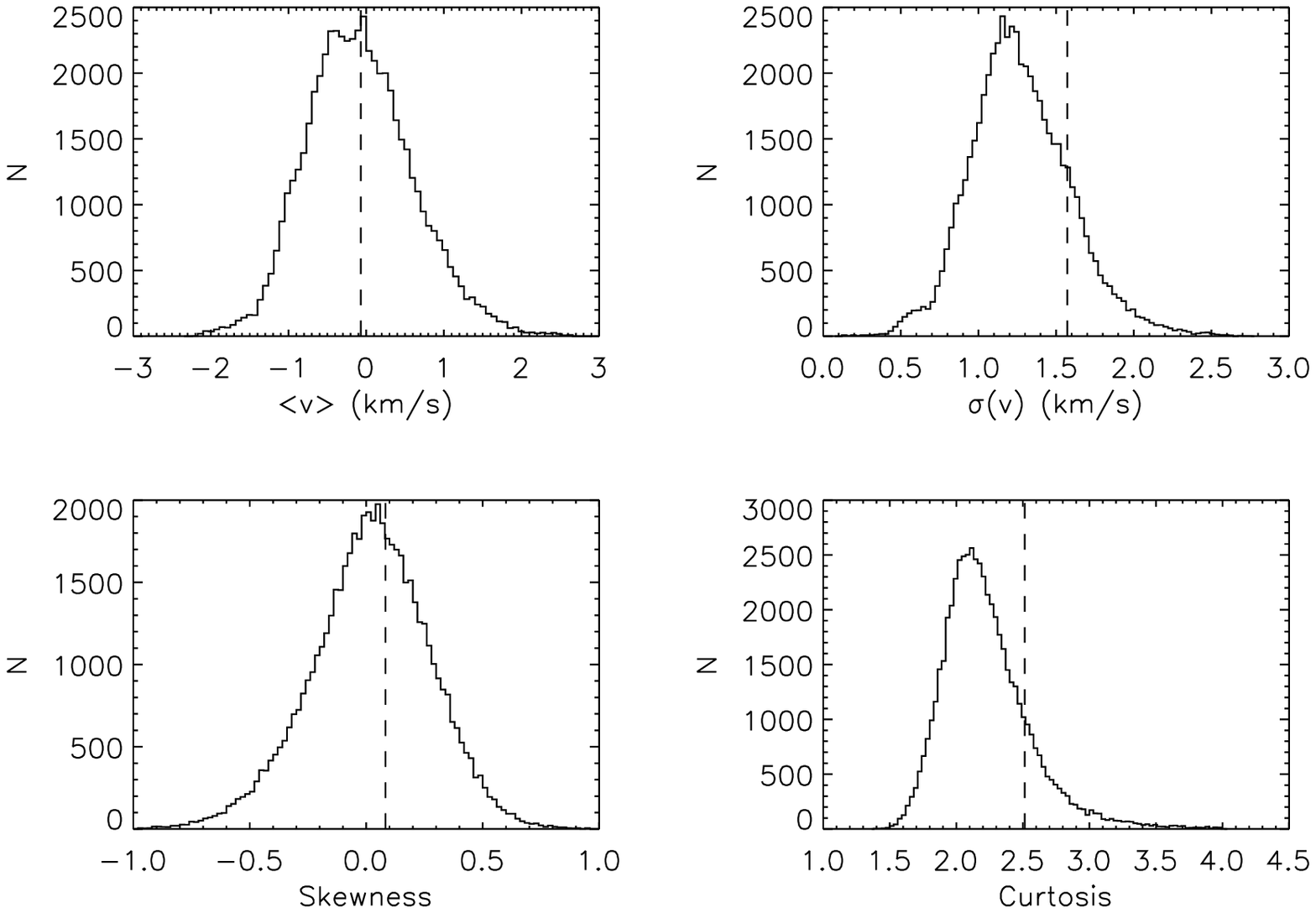}
\caption[]{}
\label{14}
\end{figure}

{\bf Figure \ref{14}:}  As in Fig.~\ref{13}, but for 
$^{12}$CO, J=1$\rightarrow$0, spectra. The main difference
from the $^{13}$CO spectra is the fact that the values of 
the kurtosis are a bit smaller.
This is due to the fact that some spectra are saturated 
and therefore their maxima are flattened. The same 
is apparent for the mean spectrum, 
whose value of kurtosis is less than 2.6.\\



\begin{figure}
\centering
\leavevmode
\epsfxsize=0.5
\columnwidth
\epsfbox{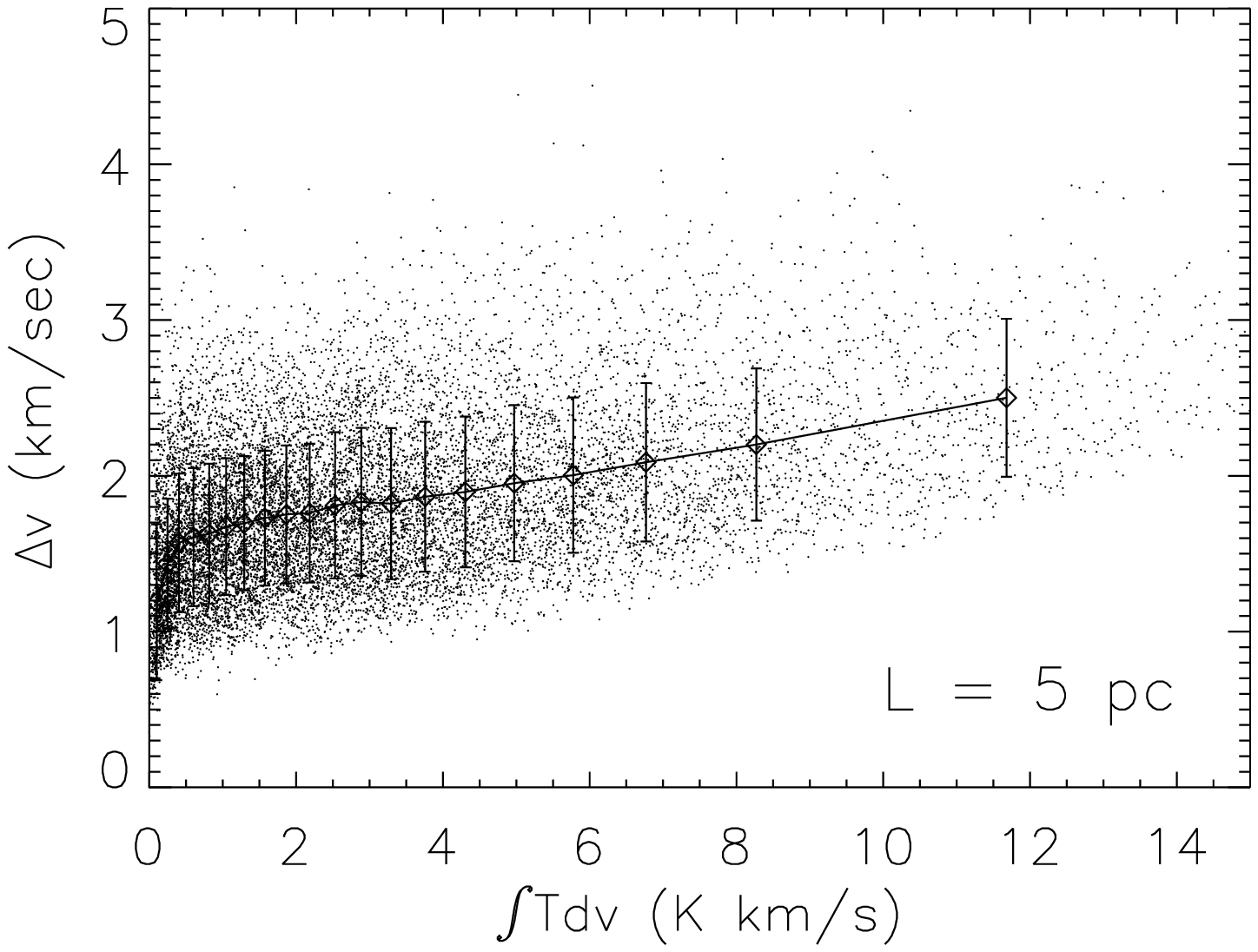}
\epsfxsize=0.5
\columnwidth
\epsfbox{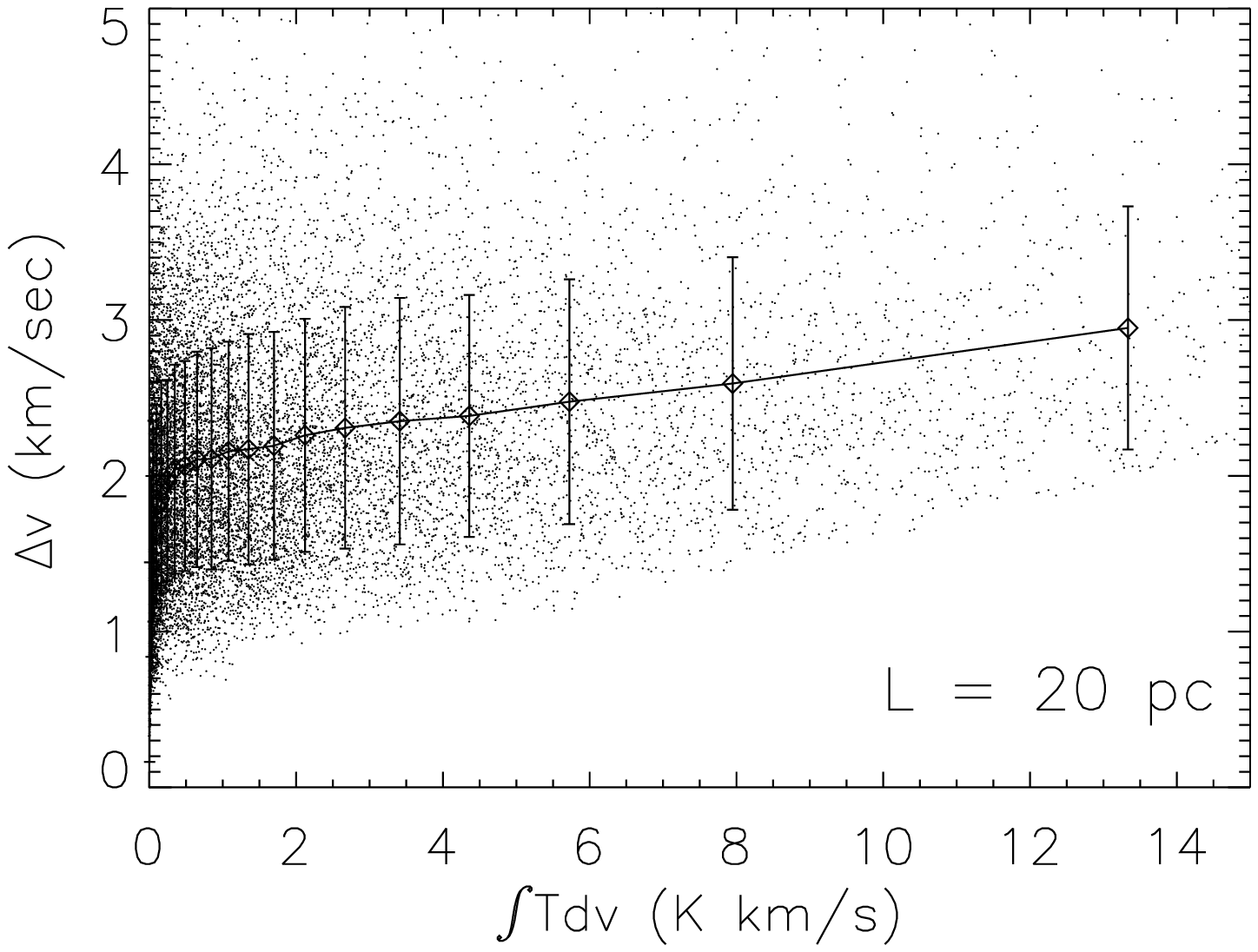}
\caption[]{}
\label{17}
\end{figure}

{\bf Figure \ref{17}:}  Equivalent width of $^{13}$CO, 
J=1$\rightarrow$0, spectra, versus integrated temperature, 
for 5pc models (left), and 20pc models 
(right). Only a small fraction of the points 
have been plotted. The continuous line connects 
the mean values in intervals with equal 
number of points. The error bars show the 
$1-\sigma$ dispersion around the mean equivalent width. \\

\begin{figure}
\centering
\leavevmode
\epsfxsize=1.0
\columnwidth
\epsfbox{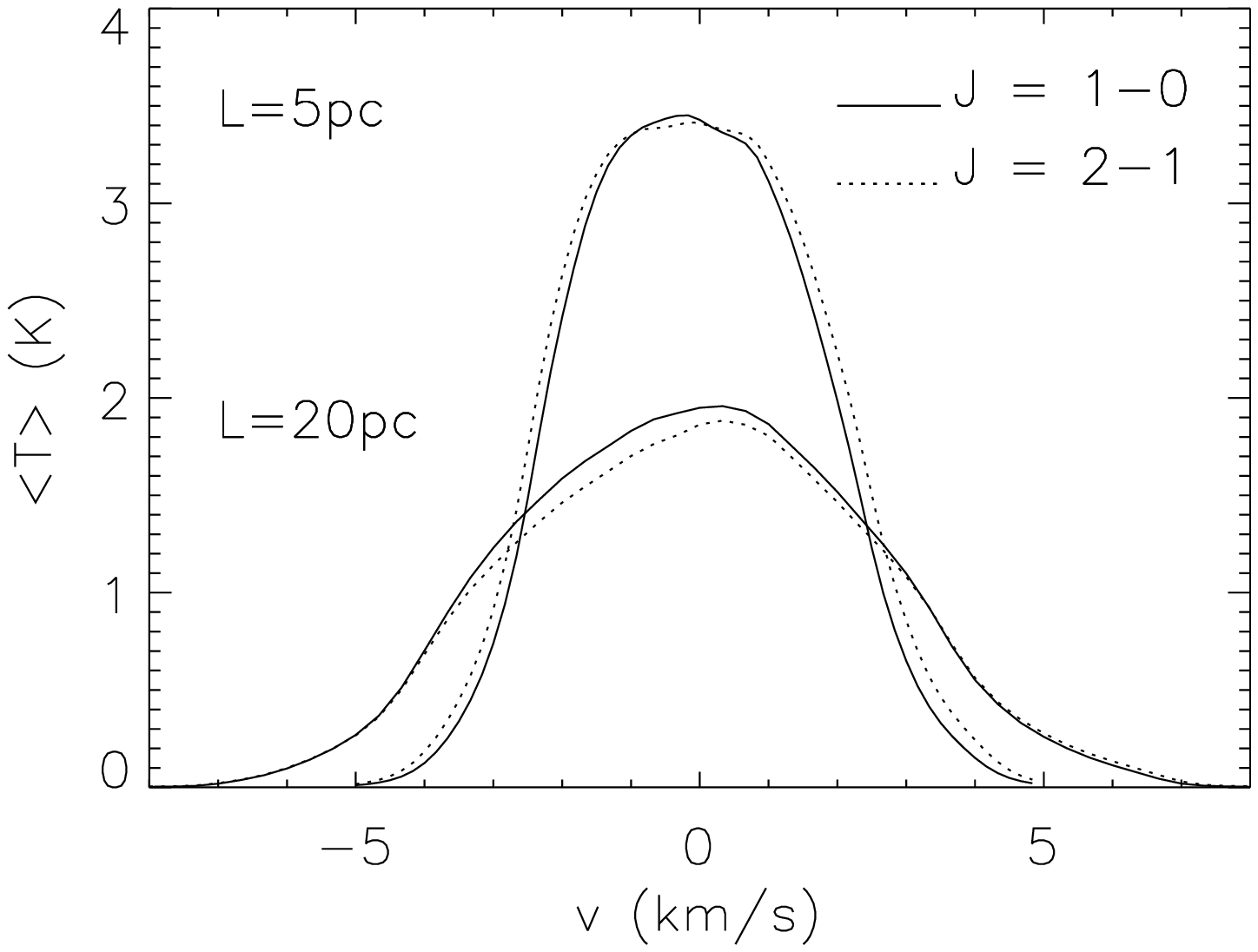}
\caption[]{}
\label{18}
\end{figure}

{\bf Figure \ref{18}:} Average $^{12}$CO J=1$\rightarrow$0
(continuous line) and J=2$\rightarrow$1 (dotted line)
spectra, from the 5 pc and the 20 pc models. 
The J=2$\rightarrow$1 temperature is divided by
$R_{CO}(2-1/1-0)=0.62\pm0.08$, that is the line ratio 
found by Falgarone \& Phillips (1996). \\

\begin{figure}
\centering
\leavevmode
\epsfxsize=0.5
\columnwidth
\epsfbox{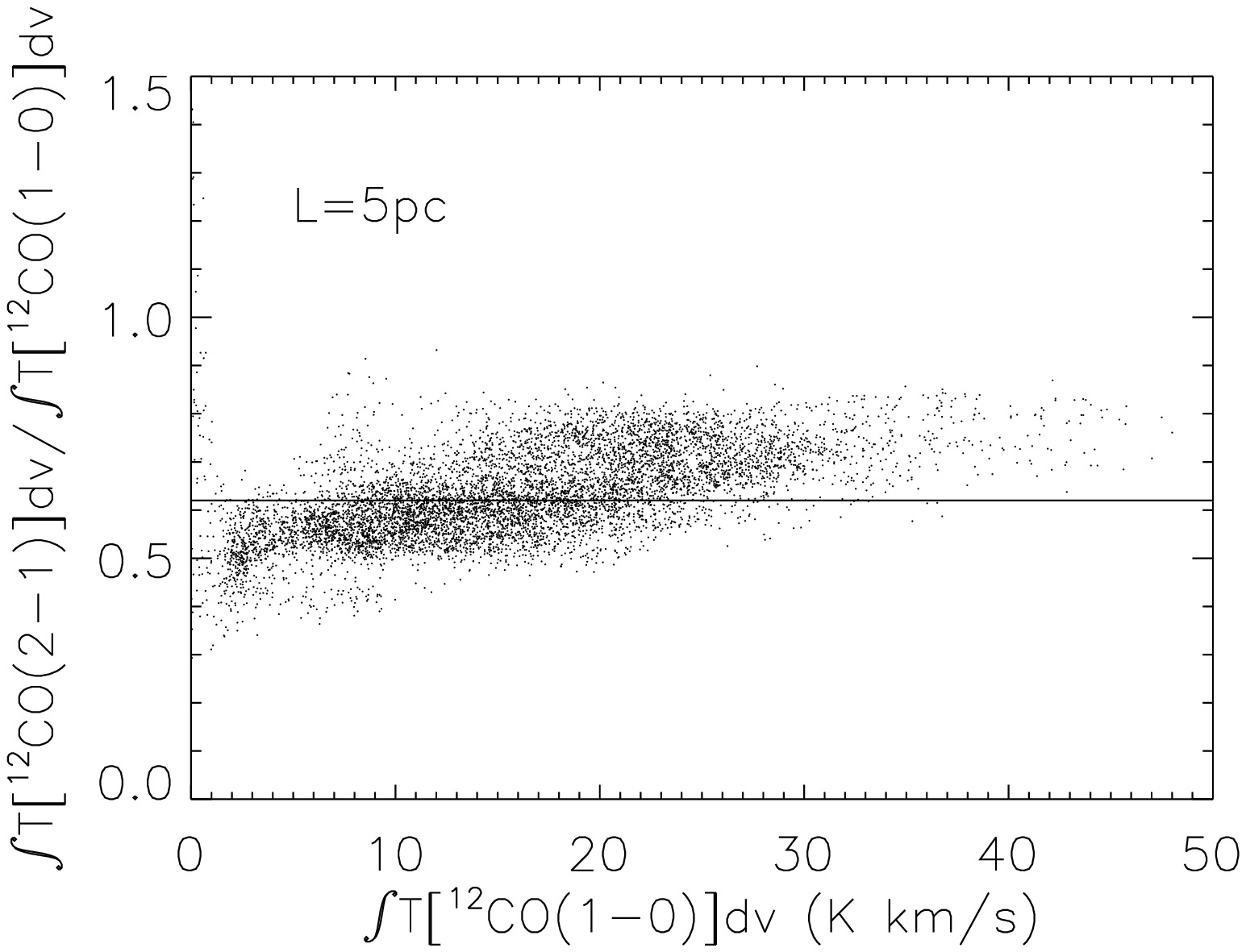}
\epsfxsize=0.5
\columnwidth
\epsfbox{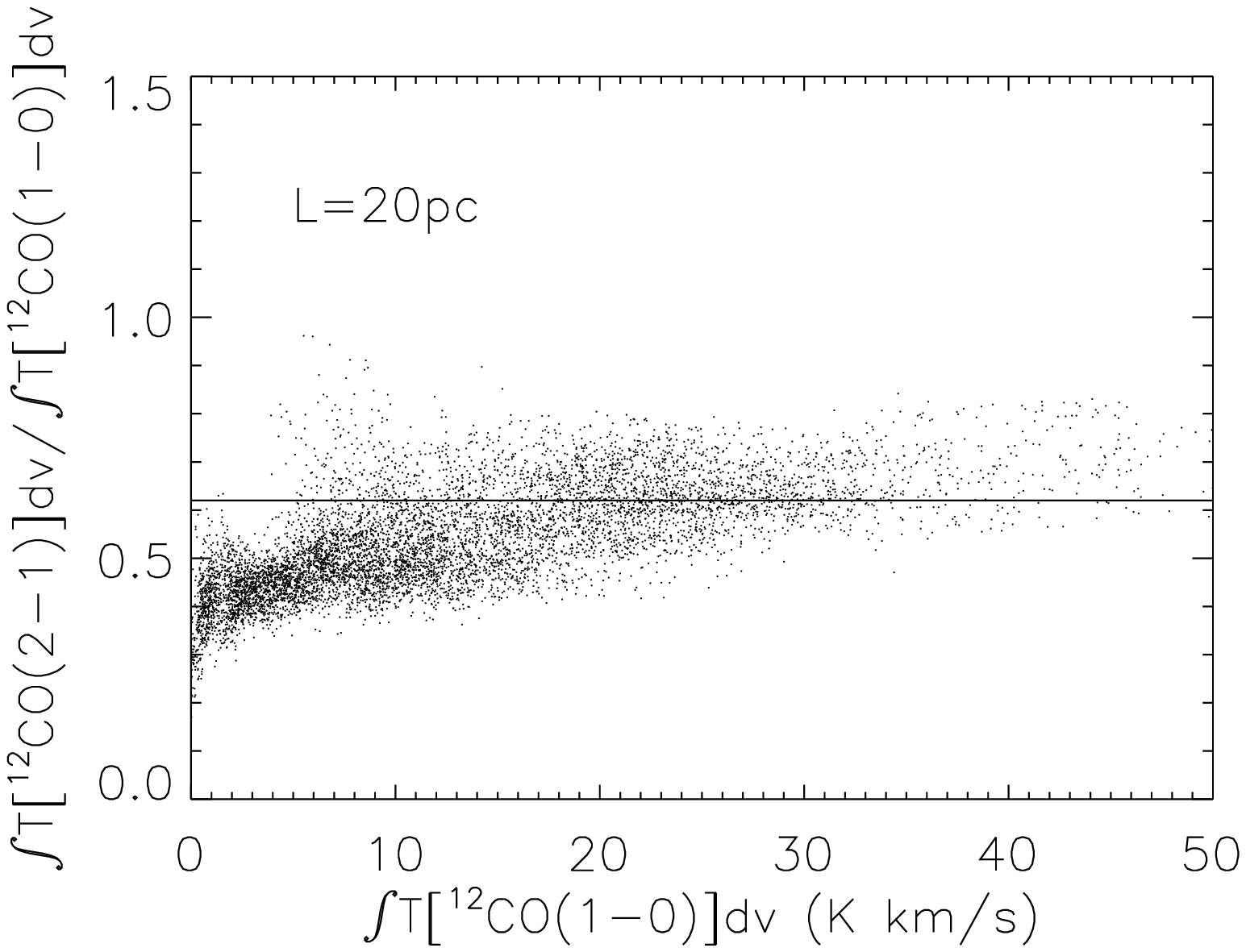}
\caption[]{}
\label{19}
\end{figure}

{\bf Figure \ref{19}:}  Ratio of velocity integrated 
$^{12}$CO J=2$\rightarrow$1 and J=1$\rightarrow$0 
temperature from single lines--of--sight. The continuous line
is the value found by Falgarone \& Phillips (1996). \\

\begin{figure}
\centering
\leavevmode
\epsfxsize=0.5
\columnwidth
\epsfbox{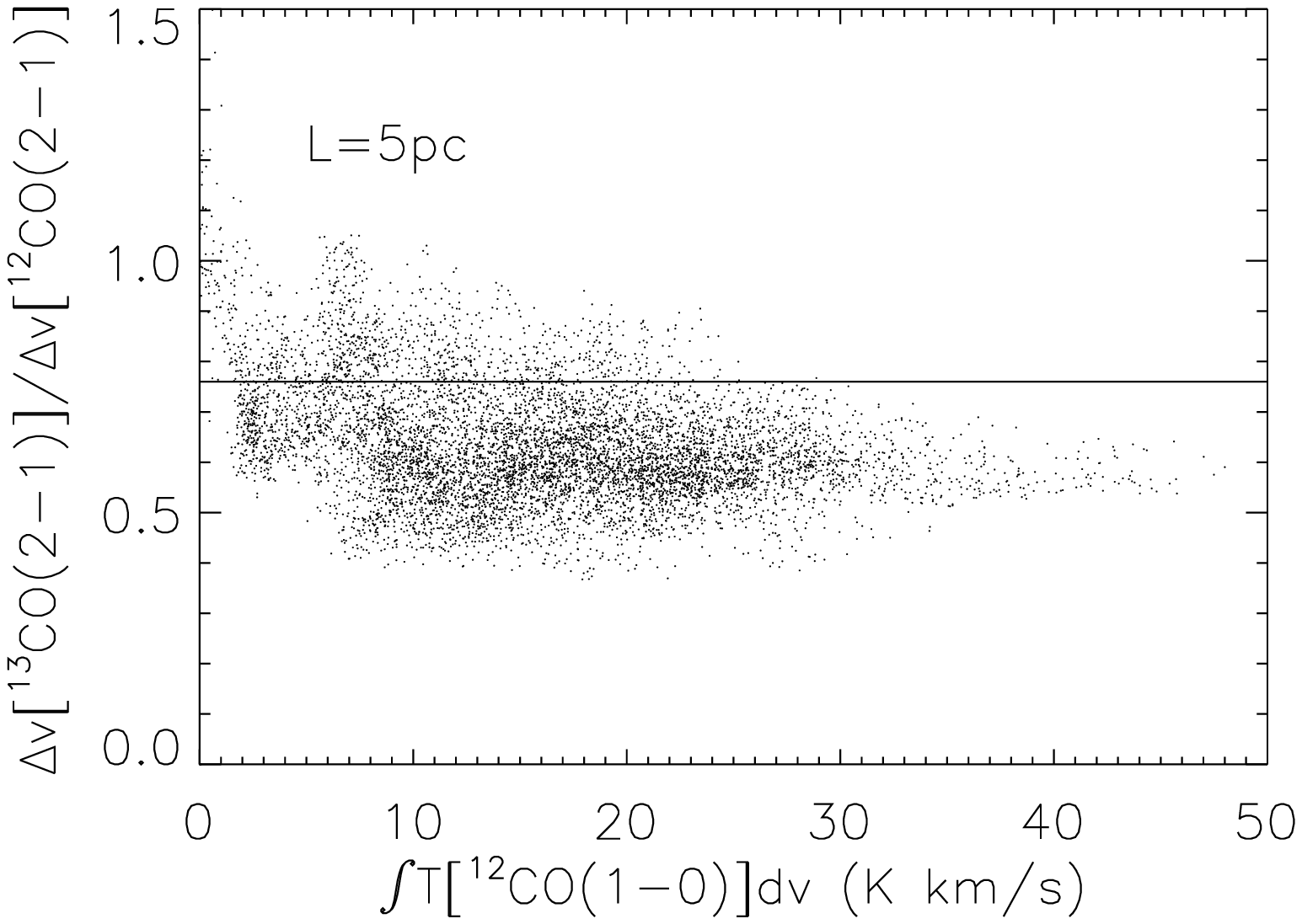}
\epsfxsize=0.5
\columnwidth
\epsfbox{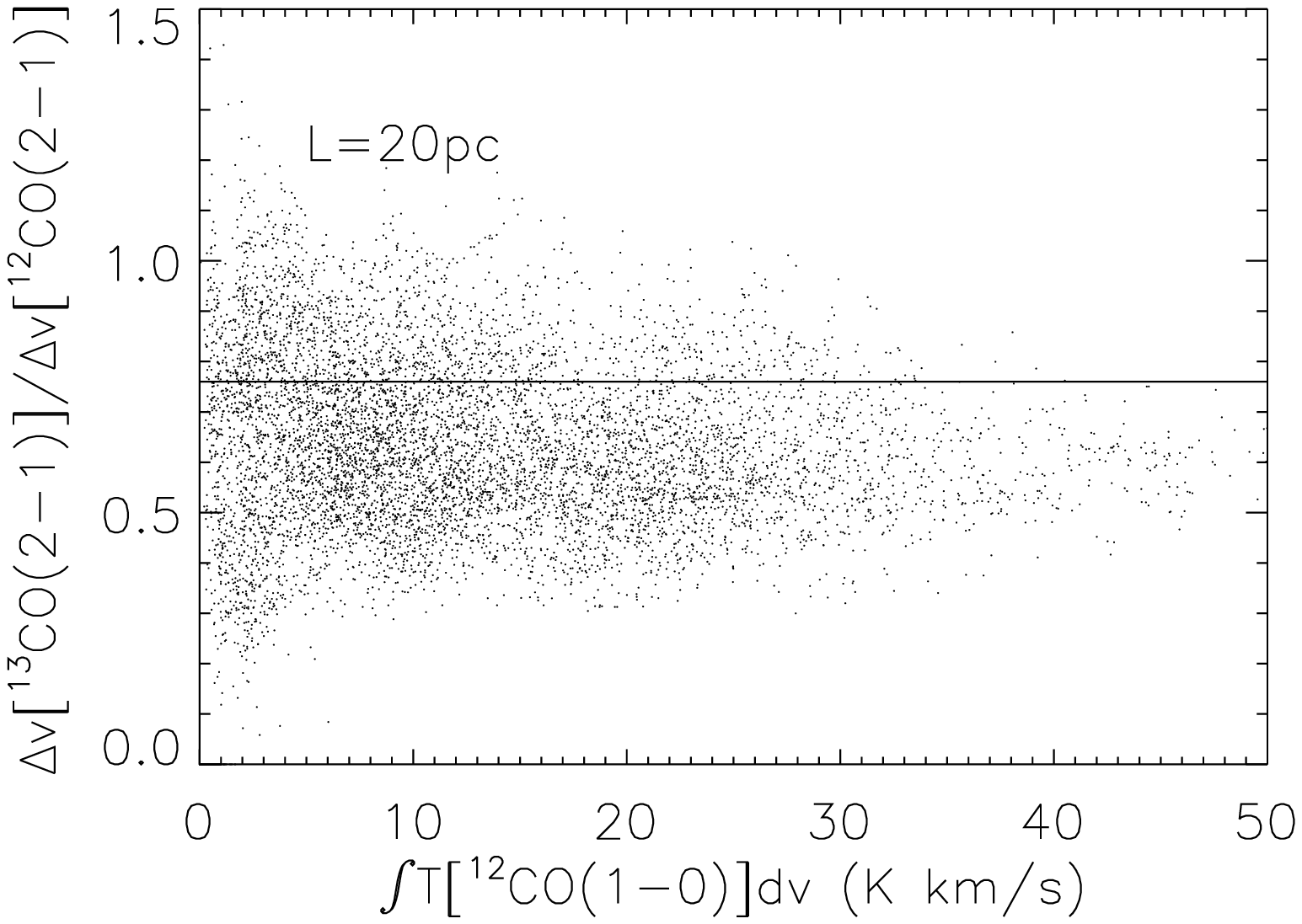}
\caption[]{}
\label{20}
\end{figure}

{\bf Figure \ref{20}:}  Ratio of $^{12}$CO J=2$\rightarrow$1 to
$^{13}$CO J=2$\rightarrow$1 line--width, versus $^{12}$CO 
J=1$\rightarrow$0 integrated temperature. The continuous horizontal 
line is the value found by Falgarone \& Phillips (1996). \\

\clearpage
\begin{table}
\begin{tabular}{|c|c|c|c|c|c|}
\hline
Model  & L (pc) & t (t$_{dyn}$) & $<n>$ (cm$^{-3}$) &  $\sigma_v$ (km/s) & T (K) \\
\hline\hline
$A_{1}$  & 5.0    &  1.0    &   400   &   1.7  &  10 \\
\hline
$A_{2}$  & 20.0   &  1.0    &   100   &   3.4  &  10 \\
\hline\hline
\end{tabular}
\caption{}
\end{table}


\clearpage




\clearpage

\clearpage

\clearpage

\clearpage

\clearpage

\clearpage

\clearpage

\clearpage

\clearpage

\clearpage

\clearpage

\end{document}